\documentclass[useAMS,usenatbib]{mn2e}

\usepackage{graphicx}
\usepackage{natbib}
\usepackage{graphicx}
\usepackage{savesym}
\usepackage{listings}
\usepackage[intlimits]{amsmath}
\savesymbol{iint}
\usepackage{txfonts}
\usepackage{bm}
\usepackage{txfonts}
\usepackage{bm}
\restoresymbol{TXF}{iint}
\usepackage{amssymb}
\usepackage{color}
\usepackage{breqn}
\usepackage{comment}
\newcommand{\comm}[1]{}

\def\green#1 {\textcolor{green}{#1}\ }   
\def\blue#1 {\textcolor{blue}{#1}\ }   

\def\be{\begin{equation}}
\def\ee{\end{equation}}
\def\ba{\begin{eqnarray}}
\def\ea{\end{eqnarray}}
\def\go{\mathrel{\raise.3ex\hbox{$>$}\mkern-14mu
             \lower0.6ex\hbox{$\sim$}}}
\def\lo{\mathrel{\raise.3ex\hbox{$<$}\mkern-14mu
             \lower0.6ex\hbox{$\sim$}}}

\title[Resistive Tearing Instability in E-MHD]{Resistive Tearing Instability in Electron-MHD: Application to Neutron Star Crusts}
\author[K.N. Gourgouliatos \& R. Hollerbach]{{Konstantinos N. Gourgouliatos\thanks{Email: K.N.Gourgouliatos@leeds.ac.uk}$^{1}$ \& Rainer Hollerbach$^{1}$}\vspace{0.4cm}\\
\parbox{\textwidth}{$^{1}$Department of Applied Mathematics, University of Leeds, Leeds LS2 9JT , UK}}

\begin{document}

 \date{Accepted -. Received -; in original form -}
\pagerange{\pageref{firstpage}--\pageref{lastpage}} \pubyear{-}
\maketitle
\label{firstpage}

\begin{abstract}
We study a resistive tearing instability developing in a system evolving through the combined effect of Hall drift in the Electron-MHD limit and Ohmic dissipation. We explore first the exponential growth of the instability in the linear case and we find the fastest growing mode, the corresponding eigenvalues and dispersion relation. The instability growth rate scales as  $\gamma \propto B^{2/3} \sigma^{-1/3}$ where $B$ is the magnetic field and $\sigma$ the electrical conductivity. We confirm the development of the tearing resistive instability in the fully non-linear case, in a plane parallel configuration where the magnetic field polarity reverses, through simulations of systems initiating in Hall equilibrium with some superimposed perturbation. Following a transient phase, during which  there is some minor rearrangement of the magnetic field, the perturbation grows exponentially. Once the instability is fully developed the magnetic field forms the characteristic islands and X-type reconnection points, where Ohmic decay is enhanced. We discuss the implications of this instability for the local magnetic field evolution in neutron stars' crusts, proposing that it can contribute to heating near the surface of the star, as suggested by models of magnetar post-burst cooling. In particular, we find that a current sheet a few meters thick, covering as little as $1\%$ of the total surface can provide $10^{42}~$erg in thermal energy within a few days. We briefly discuss applications of this instability in other systems where the Hall effect operates such as protoplanetary discs and space plasmas.
\end{abstract}

\begin{keywords}
stars: neutron, magnetars, methods: numerical, MHD, magnetic fields
\end{keywords}

\maketitle

\section{Introduction}

A plethora of observations of strongly magnetised neutron stars  \citep{Olausen:2014} has revealed that their temperatures are higher than what conventional cooling of a hot proto-neutron star suggests. A solution to this puzzle is that the extra thermal energy needed for these systems is provided by the Ohmic decay of their magnetic energy reservoir \citep{Pons:2007}. However, given the high conductivity of a neutron star crust, the rate of Ohmic decay is expected to be slow and the conversion of magnetic energy to heat inefficient. 
This has led to the idea that the Hall effect may be able to accelerate magnetic field decay, as the Hall timescale is inversely proportional to the intensity of the magnetic field. This acceleration can only be done in an indirect way, as the Hall effect conserves magnetic field energy. 

Several paths have been proposed in this direction. \cite{Goldreich:1992} suggested that the Hall effect may lead to the formation of smaller scale structure through cascades, which have reduced Ohmic decay times, a  result that has been followed up by numerical studies exploring Hall-induced turbulence \citep{Biskamp:1996, Wareing:2009b, Wareing:2010}. Another possibility is the development of instability of a state previously being in Hall equilibrium leading to smaller structure formation \citep{Rheinhardt:2002, Rheinhardt:2004, Pons:2010}. Recent work of \cite{Wood:2014} found a family of exact solutions for the density-shear instability in electron-MHD, requiring a covarying magnetic field and electron number density, a result that was studied numerically in detail by \cite{Gourgouliatos:2015b}. Apart from instabilities and cascades, secular Hall evolution has been explored: \cite{Vainshtein:2000} studied the effect of the sharp drop of electron number in the crust, finding that  the magnetic field evolution is described by a Burger's type equation, leading to the formation of shocks in the form of current sheets decaying on a Hall timescale rather than the slower Ohmic, and applied to the evolution of a toroidal field in an axially symmetric system by \cite{Reisenegger:2007}. Once the poloidal field is included \citep{Hollerbach:2002, Hollerbach:2004} the formation of current sheets is followed by an oscillatory behaviour. The consensus of axially symmetric crustal simulations, exploring a broad range of initial conditions \citep{Pons:2009, Kojima:2012,Vigano:2012,Gourgouliatos:2014a}, has concluded that the Hall effect drastically changes the structure of the magnetic field, whereas later, Hall evolution saturates \citep{Gourgouliatos:2014b}. 

An intrinsic drawback of global neutron star simulations is the fact that they under-resolve current sheets. Current sheets form both in the uniform electron density case \citep{Wareing:2010} and even more efficiently in the presence of an electron density gradient \citep{Vainshtein:2000, Vigano:2012}. Furthermore, they are likely to appear near the surface of the crust, as the available electric charges decrease dramatically from the solid crust to the plasma magnetosphere. In the latter case, a usual assumption made in simulations is that the external magnetic field is a vacuum potential field which leads to boundary effects by matching the two configurations \citep{Wood:2014}.  

In their seminal paper \cite{Furth:1963} studied finite-resistivity instabilities of a sheet pinch finding the so-called tearing instability {\it ``a long-wave `tearing' mode, corresponding to a breakup of the layer along the current flow lines"}. Linear analysis of the MHD system yields an exponential growth rate $\gamma_{T} \sim \tau_{O}^{-3/5}\tau_{A}^{-2/5}$, where $\tau_{O}$ and $\tau_{A}$ are the resistive and Alfv\`en times respectively, while in the non-linear phase the growth becomes algebraic \citep{Rutherford:1973}. Several applications of the tearing instability have been considered in astrophysical contexts. \cite{Rosenbluth:1967} studied resistive instabilities in magnetospheric tails. \cite{Priest:1985} presented various applications of the tearing instability in relation to current sheets developing in solar and space plasmas. The tearing instability is considered to be an efficient mechanism for powering solar flares and accelerating particles therein \citep{Sturrock:1966, Somov:1989, Aschwanden:2002}. Recent numerical simulations by \cite{Landi:2015, DelZanna:2016} in general astrophysical contexts have demonstrated the development of the tearing instability in the limit of very high conductivity for appropriately thin current layers. Other applications have focused on pulsar magnetospheres, where numerical simulations agree on the presence of current sheets, either confined to the equatorial plane as is the case in axially symmetric systems \citep{Contopoulos:1999, Komissarov:2006}, or with more complicated geometries for the case of inclined systems \citep{Spitkovsky:2006, Kalapotharakos:2009}. In depth study of the current sheets of pulsar magnetospheres by \cite{Uzdensky:2014}, showed that they are susceptible to the tearing mode instability leading to the formation of plasmoids with the eventual emission of high energy radiation and non-thermal particles \citep{Sironi:2014}.  The tearing instability has also been studied in the context of Relativistic MHD considering applications to magnetar flares and jets through explosive reconnection  \citep{ Komissarov:2007, Elenbaas:2016, Barkov:2016}. 

Motivated by the omnipresence of the tearing instability in current sheets and their formation in neutron star crusts through the Hall effect, we study its development and impact. We explore the evolution of the magnetic field in a configuration where the tangential component changes direction by $180^{\circ}$ within a thin layer, allowing for some finite resistivity, in the inertialess electron-MHD formulation. We show, through linear and non-linear calculations, that the tearing mode instability naturally appears and enhances the decay of the magnetic field. 

We note that the term Hall evolution (or drift) has the meaning of Electron-MHD when used to describe the evolution of the magnetic field in the crust of neutron stars.  There, only electrons are allowed to moved through a solid crystal lattice consisting of positively charged ions \citep{Jones:1988}. In principle, Hall evolution can accommodate for the motion of more than one charged species whereas Electron-MHD refers to systems where only electrons move, making the latter a special case of the former. In this paper the term Hall-MHD is used in the limit of Electron-MHD.

The plan of the paper is as follows: In Section 2 we formulate the equations of Electron-MHD. We solve these equations in the linear and non-linear regime in Section 3. We discuss the properties of the instability and compare it with the conventional tearing instability in Section 4. We discuss the application of the tearing instability in neutron stars and other astrophysical systems in Section 5. We conclude in Section 6.

\section{Electron-MHD formulation in neutron star crusts}

The crust is the outer layer of the neutron star with thickness of about 1km. The density at the base is $10^{14}$g cm$^{-3}$ and $10^{9}$g cm$^{-3}$ at the surface. It can be approximated to good accuracy by a highly conducting ion Coulomb lattice with electrons having the freedom to move. Following the derivation of \cite{Goldreich:1992}, the crustal electric current must be carried by free electrons: $\bm{j}=-n_{e}e \bm{v}_{e}$, where $\bm{j}$ is the current density, $n_{e}$ the electron number density, $e$ the electron charge and $\bm{v}_{e}$ the electron velocity. Then, from Amp\`ere's law $\bm{j}=(c/4 \pi)\nabla\times \bm{B}$, where $c$ is the speed of light and $\bm{B}$ the magnetic induction and using Ohm's law $\bm{j} =\sigma\left(\bm{E}+\left(\bm{v}_{e} \times \bm{B}\right)/c\right)$ where $\bm{E}$ is the electric field and $\sigma$ the electric conductivity, we substitute into Faraday's law, yielding:
\begin{eqnarray}
\frac{\partial \bm{B}}{\partial t} = -\nabla \times \left(\frac{c}{4 \pi e n_{e}}\left(\nabla \times \bm{B}\right)\times \bm{B} +\frac{c^{2}}{4 \pi \sigma} \nabla \times \bm{B}\right)\,.
\label{HALL}
\end{eqnarray}
The first term on the right hand side of the above equation describes the evolution under the Hall effect and the second one Ohmic dissipation. Conceptually, the Hall effect can be thought of as the advection of the magnetic flux by the electron fluid. 

Contrary to usual MHD this equation does not assume that mass is displaced, as the crustal ions hold fixed positions in space, while the moving electrons are to good approximation inertialess. The Lorentz forces are balanced by the elasticity of the crust. The only physical quantity involved in the description of the system is the magnetic induction $\bm{B}$, while for instance in normal MHD one needs to solve for the plasma velocity through the momentum equation. 

It follows, from the first term on the right hand side of equation (\ref{HALL}), that a state for which the following condition is satisfied
\begin{eqnarray}
\nabla \times \left(\frac{c}{4 \pi e n_{e}} \left(\nabla \times \bm{B}\right)\times \bm{B} \right)=\bm{0}\,,
\label{EQUI}
\end{eqnarray}
corresponds to a Hall equilibrium, and will not evolve in the limit of zero resistivity \citep{Cumming:2004, Gourgouliatos:2013, Fujisawa:2014}. In the realistic case of non-zero resistivity, however, the system  will start evolving and may be pushed out of Hall equilibrium \citep{Marchant:2014}.

\section{Tearing Instability}

\subsection{Linear Theory}

Let us assume a background magnetic field with components along the $y$ and $z$ directions depending only on $x$, and a constant electron number density ($n_{e}$) and electric conductivity ($\sigma$):
\begin{eqnarray}
\bm{B}_{b}=B_{y}(x)\hat{\bm{y}} +B_{z}(x)\hat{\bm{z}}\,.
\end{eqnarray}
This magnetic field corresponds to a Hall equilibrium satisfying equation (\ref{EQUI}). Consider some perturbation $\bm{b}(x,z,y, t)=\exp\left(\gamma t+ik_{y}y+ik_{z}z\right)(b_{x}(x)\hat{\bm{x}}+b_{y}(x)\hat{\bm{y}}+b_{z}(x)\hat{\bm{z}})$; by Gauss's law it is $\nabla \cdot \bm{b}=0$, thus $b_{z}=ik_{z}^{-1}b_{x}^{\prime}-k_{y}k_{z}^{-1}b_{y}$ where prime denotes derivative with respect to $x$. Thus the perturbation becomes:
\begin{eqnarray}
\bm{b}=\exp\left(\gamma t+ik_{y}y+ik_{z}z\right)\left[b_{x}(x)\hat{\bm{x}}+b_{y}(x)\hat{\bm{y}}+\left(ik_{z}^{-1}b_{x}^{\prime}-k_{y}k_{z}^{-1}b_{y}\right)\hat{\bm{z}}\right]\,.
\end{eqnarray}
Substituting into equation (\ref{HALL}) and keeping only the linear terms in $\bm{b}$ we obtain the following equations:
\begin{dmath}
\gamma b_{x}+\frac{c}{4 \pi e n_{e}} \left[k_{z}^{2}B_{z}b_{y}-ik_{z}B_{y}^{\prime}b_{x}+k_{y} \left(i \left\{B_{z}^{\prime}b_{x}-B_{z}b_{x}^{\prime} -k_{y} k_{z}^{-1}B_{y}b_{x}^{\prime}    \right\}  + \left\{k_{z}B_{y}+k^{2}_{y}k_{z}^{-1}B_{y}+k_{y}B_{z}\right\}b_{y}  \right) \right]+\frac{c^{2}}{4 \pi \sigma}\left[\left(k_{y}^{2}+k_{z}^{2}\right) b_{x}-b_{x}^{\prime \prime} \right]=0\,, 
\end{dmath}
\begin{dmath}
\gamma b_{y}+\frac{c}{4 \pi e n_{e}} \left[-k_{z}^{2}B_{z}b_{x}-B_{z}^{\prime \prime}b_{x}+B_{z}b_{x}^{\prime \prime} +k_{y}\left(i\left\{ k_{y} k_{z}^{-1} \left(B_{y}b_{y}\right) ^{\prime} +\left(B_{z}b_{y}\right)^{\prime}\right\}-k_{z}B_{y}b_{x}+k_{z}^{-1} \left(B_{y} b_{x}^{\prime}\right)^{\prime} \right)\right] +\frac{c^{2}}{4 \pi \sigma}\left[\left(k_{y}^{2}+k_{z}^{2}\right)b_{y}-b_{y}^{\prime \prime}\right]=0\,.
\end{dmath}
We first explore numerically the eigenvalue problem. To model the structure of a current sheet we have chosen the following profile for the background field:
\begin{eqnarray}
\begin{aligned}
&B_{y}=B_{y,0} {\rm sech} \left(\frac{x}{x_{0}}\right)\,, \\
&B_{z}=B_{z,0}\tanh\left(\frac{x}{x_{0}}\right)\,,\label{BG2}
\end{aligned}
\end{eqnarray}
assuming $x_{0}>0$ and $B_{z,0}>0$. The field becomes uniform along the $z$ direction for $|x|~\gg x_{0}$. A choice of amplitudes $B_{y,0}=\pm B_{z,0}$ corresponds to a Bloch wall \citep{Bloch:1932}: a magnetic field that changes direction from the $-z$ to the $+z$ keeping its magnitude constant, within a layer of thickness scaling with $x_{0}$ centred at $x=0$. This case has been of particular interest in MHD simulations as it is a force-free magnetic field \citep{Low:1973}, making it an appropriate choice for studies of resistive instabilities. However, this is an unnecessary constraint for Electron-MHD studies as any choice of $B_{y,0}$ amplitude is a Hall equilibrium since equation (\ref{EQUI}) is identically satisfied. 

The configuration extends from $-x_{b}$ to $x_{b}$. We impose vacuum boundary conditions at $x$ boundaries, demanding that no currents exist outside the domain. We ensure that $x_{0}$ is sufficiently smaller than $x_{b}$ for the results to be physically meaningful, and the background field $\bm{B}_{b}$ is essentially uniform and current free close to the boundaries. Demanding vacuum boundary conditions  $\nabla \times \bm{b}=0$ for these equations at $|x|~>x_{b}$ we obtain the following equations: $b_{x}^{\prime} \pm k_{z}^{2}\left(k_{y}^{2}+k_{z}^{2}\right)^{-1/2}b_{x} + ik_{y}b_{y}=0$ and $ik_{y}b_{x}^{\prime}-\left(k_{y}^{2}+k_{z}^{2}\right)b_{y}=0$.  We consider an appropriate system of units so that $x_{b}=1$, $cB_{z,0}/(4 \pi e n_{e})=1$ where the growth rate is measured in units of inverse Hall times $\tau_{H}=4 \pi e n_{e} x_{b}^{2}/(c B_{z,0})$, with the characteristic Ohmic timescale being $\tau_{O}=4 \pi \sigma x_{b}^{2}/c^{2}$.  We define the Hall parameter $R_{H}=\sigma B_{z,0}/(c e n_{e})$, the ratio of the Ohmic timescale over the Hall timescale. Larger $R_{H}$ correspond to systems where the Hall effect dominates. In the systems we studied we have set $B_{z,0}=1$, combining it with $B_{y,0}=0$ and $B_{y,0}=1$. We have varied the thickness of the current sheet from $x_{0}=0.1$ to $x_{0}=0.5$, and the Hall parameter from $R_{H}=100$ to $R_{H}=2000$, by changing the conductivity, see Tables \ref{Table:1} and \ref{Table:2} for the range of parameters used. Then we solve the linear problem to determine the fastest growing eigenmodes of $b_{x}$ and $b_{y}$ and the corresponding eigenvalues. We do so by discretising the system of ordinary differential equations (5) and (6) and constructing the relevant matrix, whose eigenvalues allow us to determine $\gamma$ and the eigenmodes. We have implemented this using a finite difference and a spectral calculation finding identical results. We used up to a $1000$ Chebyshev polynomial expansion in the highest $R_{H}$ and thinner $x_{0}$ simulated for convergence, see chapter 7 of \cite{Boyd:2001}. The results were tested against the finite difference calculation to ensure their validity. 

\begin{figure}
\includegraphics[width=\columnwidth]{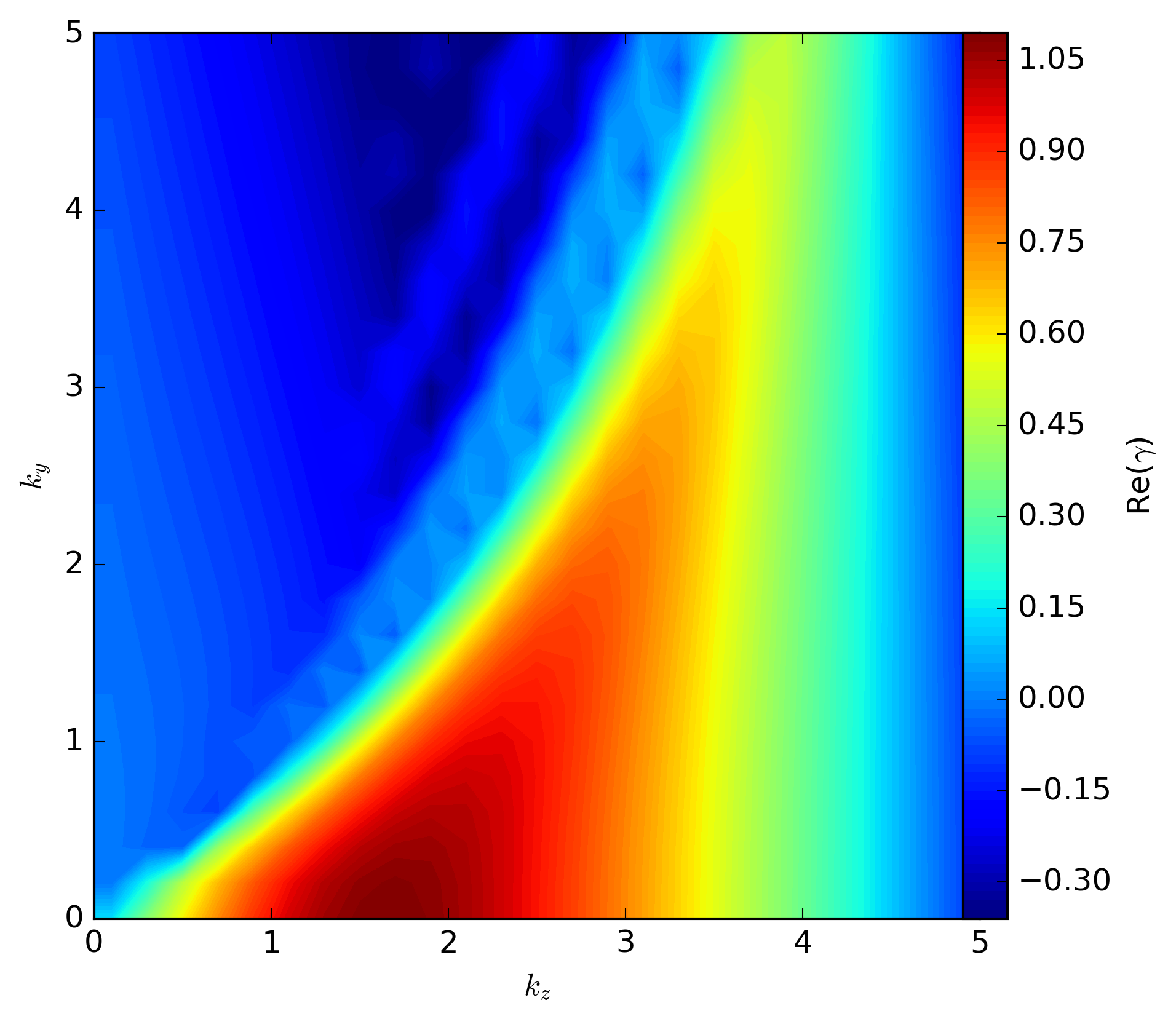}
\caption{Contour plot of the real part of the eigenvalues for a range of wavenumbers $(k_{z}, k_{y})$, using the Z02-4 profile. We find that the maximum eigenvalue occurs for $k_{y}=0$. }
\label{Fig:-1}
\end{figure}

\begin{figure}
\includegraphics[width=\columnwidth]{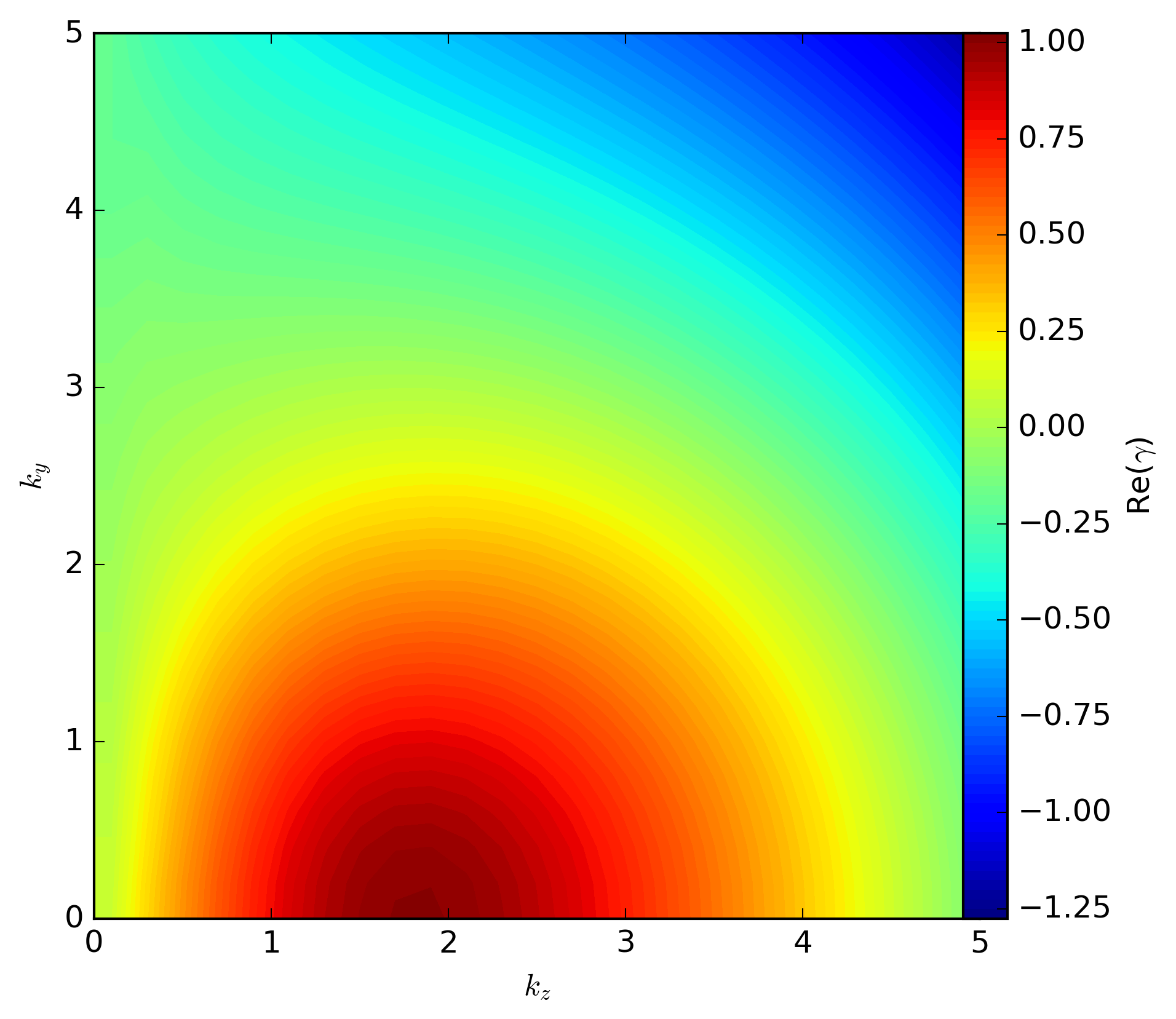}
\caption{Contour plot of the real part of the eigenvalues for a range of wavenumbers $(k_{z}, k_{y})$, using the Y02-4 profile. We find that the maximum eigenvalue occurs for $k_{y}=0$.}
\label{Fig:0}
\end{figure}

Studying the plane parallel perturbations with $k_{y}=0$, we find that both the eigenvalues and eigenfunctions for $B_{y,0}=0$ are real, while if $B_{y,0}\neq 0$ the eigenvalues are still real but the eigenfunctions become complex indicative of phase shifting in $z$. Allowing the instability to have $k_{y}\neq0$ leads to complex eigenvalues and slower growing eigenmodes for the same background field and $R_{H}$, Figures \ref{Fig:-1} and \ref{Fig:0}. Hereafter we will focus on the $k_{y}=0$ case. 

\begin{figure}
\includegraphics[width=\columnwidth]{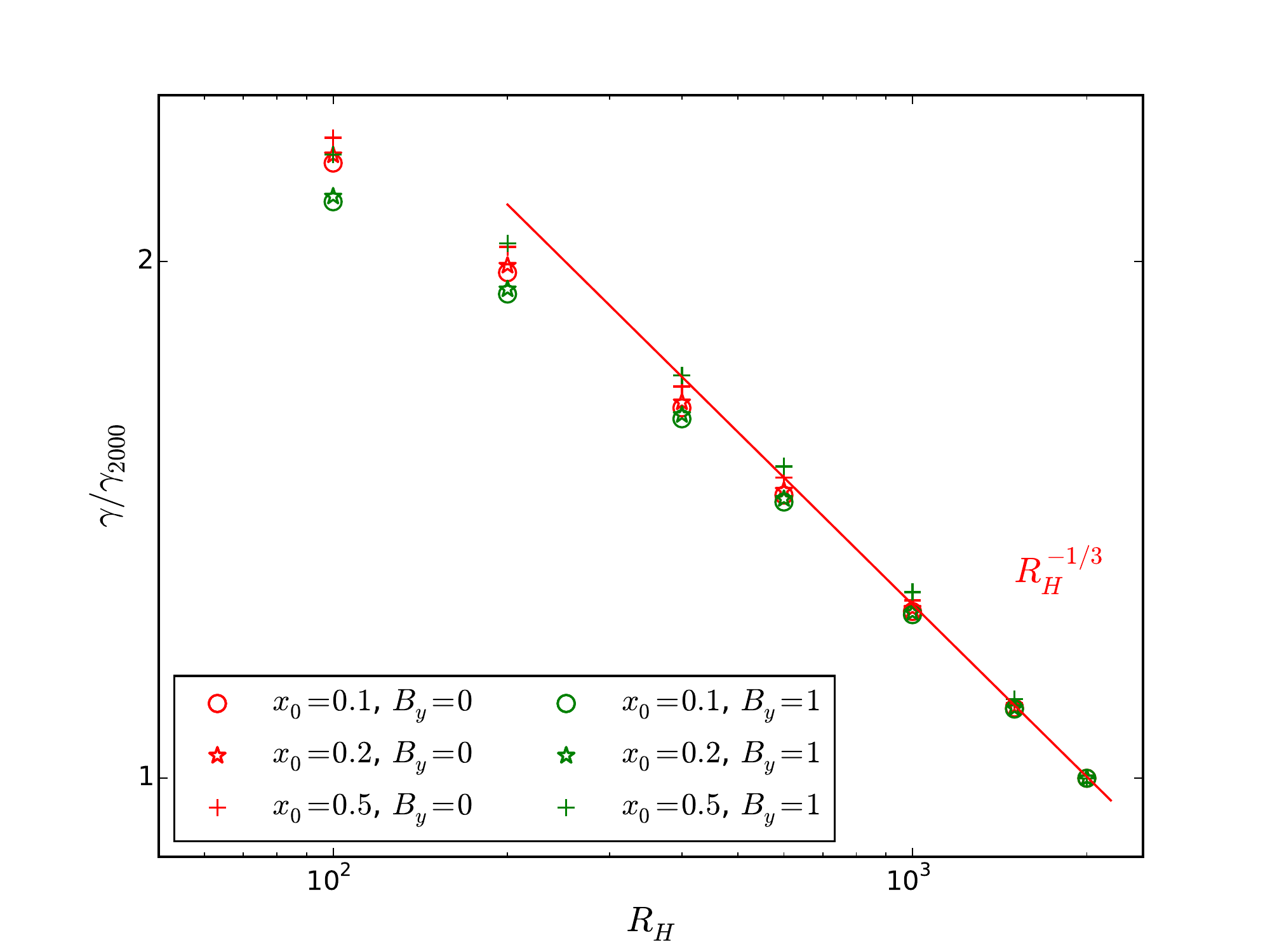}
\caption{Maximum growth rate of the tearing instability, normalised to its value at $R_{H}=2000$, versus $R_{H}$. The red crosses correspond to models Z05-1 up to Z05-20, the red stars to Z02-1 up to Z02-20, the red circles to Z01-1 up to Z01-20, the green crosses correspond to models Y05-1 up to Y05-20, the green stars to Y02-1 up to Y02-20 and the green circles to Y01-1 up to Y01-20. The growth rate scales asymptotically with $R_{H}^{-1/3}$.}
\label{Fig:1}
\end{figure}

\begin{figure}
\includegraphics[width=\columnwidth]{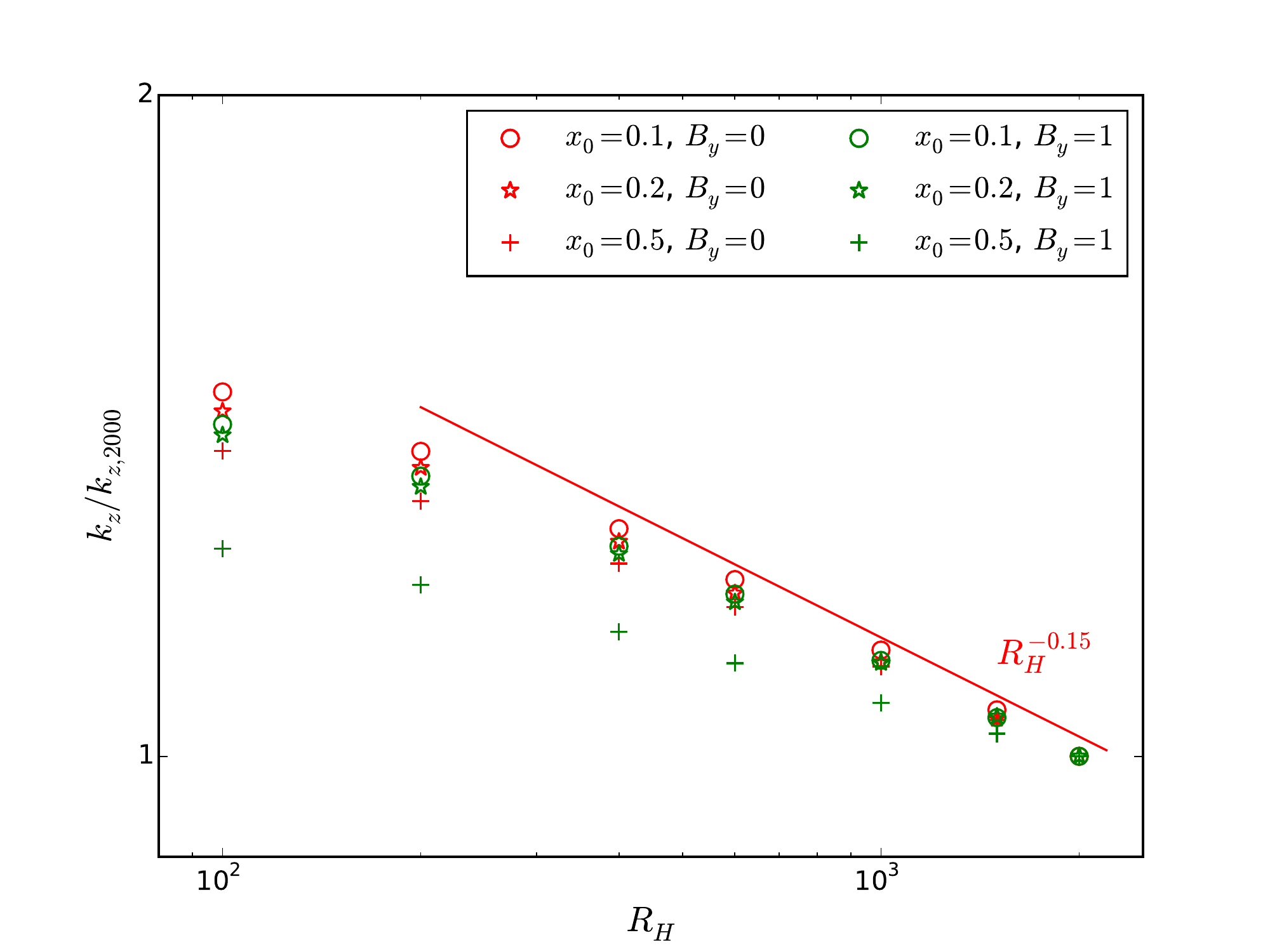}
\caption{The wave number at which the maximum growth rate occurs, normalised to its value at $R_{H}=2000$, versus $R_{H}$. There is an asysmptotic scaling with the Hall parameter $k_{z} \propto R_{H}^{-0.15}$ dependence. The symbols are the same as in Fig.~\ref{Fig:1}.}
\label{Fig:2}
\end{figure}

The maximum growth rate of the instability scales as $\gamma\propto R_{H}^{-1/3}$, Fig.~\ref{Fig:1}. The wave numbers at which the maximum growth rate occurs are plotted in Fig.~\ref{Fig:2}, and scales as $R_{H}^{-0.15}$. These scaling laws hold for narrow current sheets and high enough Hall parameters. Thus, the corresponding minimum growth timescale for the tearing instability becomes $\tau_{I}=\gamma^{-1}\propto \tau_{H}^{2/3}\tau_{O}^{1/3}$ and in terms of the physical quantities appearing $\gamma \propto B_{z,0}^{2/3}\sigma^{-1/3}$, assuming that the thickness of the reversal area remains unchanged. This quasi-stationarity assumption holds as as long as $\tau_{I}\ll \tau_{O}$ which corresponds to $R_{H}^{2/3}\gg 1$.

The maximum growth rate and the corresponding wave number are higher for thinner current sheets, with the growth rate scaling approximately as $x_{0}^{-2}$ and the wave number as $x_{0}^{-1}$. Thus, the growth time of the tearing instability $\tau_{I}$ in the linear regime can be summarised in the following expression:
\begin{eqnarray}
\tau_{I}=\frac{\tau_{H} (10 x_{0}/x_{b})^{2}(R_{H}/100)^{1/3}}{\gamma_{Z01-1}}\,,
\label{INST}
\end{eqnarray}
where $\gamma_{Z01-1}$ is the dimensionless growth rate of a system with $R_{H}=100$ and $x_{0}=0.1x_{b}$, note that $\tau_{I}$ is measured in natural units and is not rescaled. 

\begin{figure}
\includegraphics[width=\columnwidth]{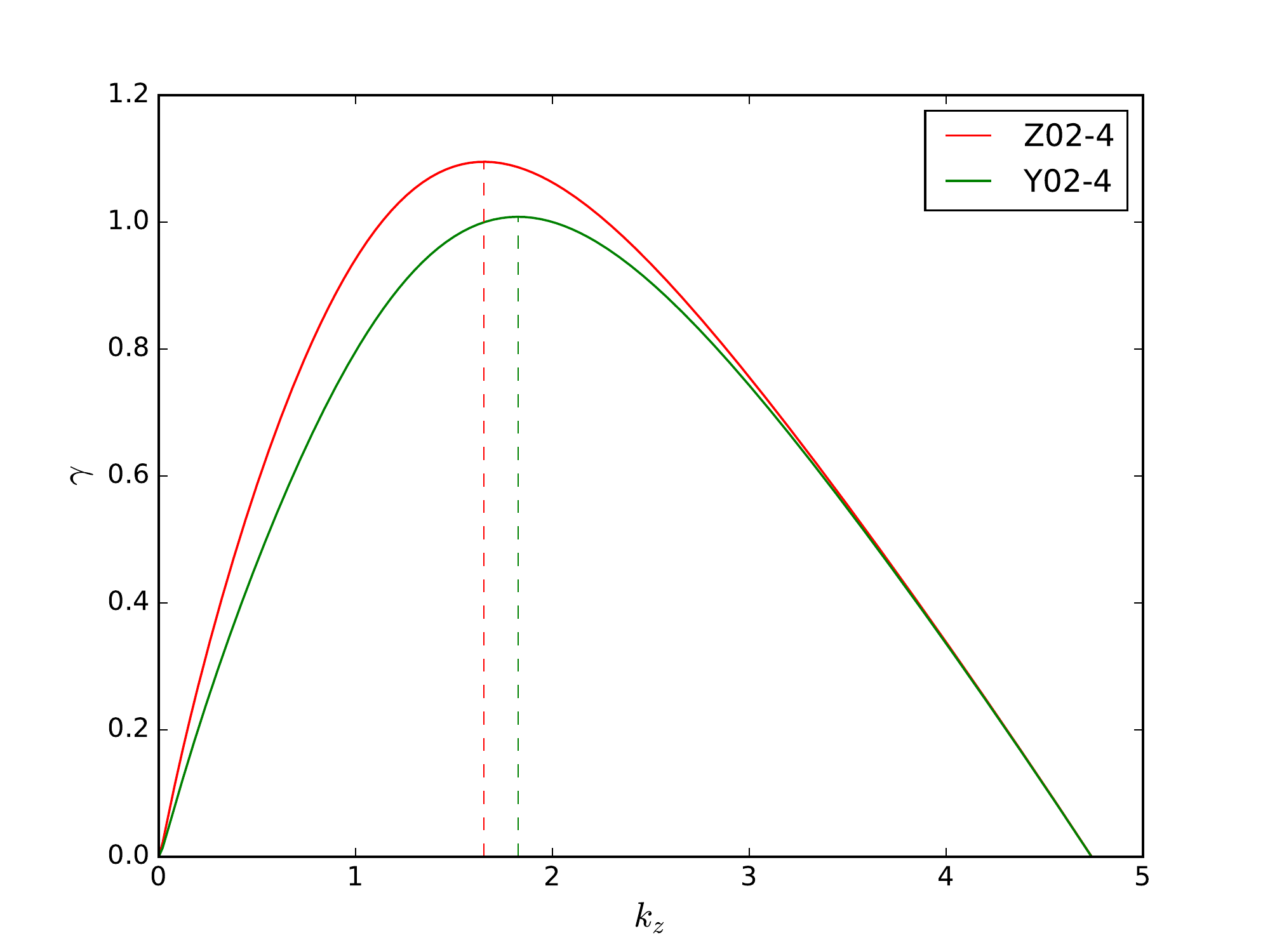}
\caption{The growth rate versus the wave number for Z02-4 (red) and Y02-4 (green), see Tables \ref{Table:1} and {Table:2}. Both of them have $x_{0}=0.2$ and $R_{H}=400$, whereas the Z02-4 has $B_{y,0}=0$ and the Y02-4 has $B_{y,0}=1$. The case with $B_{y,0}=1$ has a smaller growth rate and the maximum is pushed towards a higher wave numbers. }
\label{Fig:3}
\end{figure}

\begin{figure}
\includegraphics[width=\columnwidth]{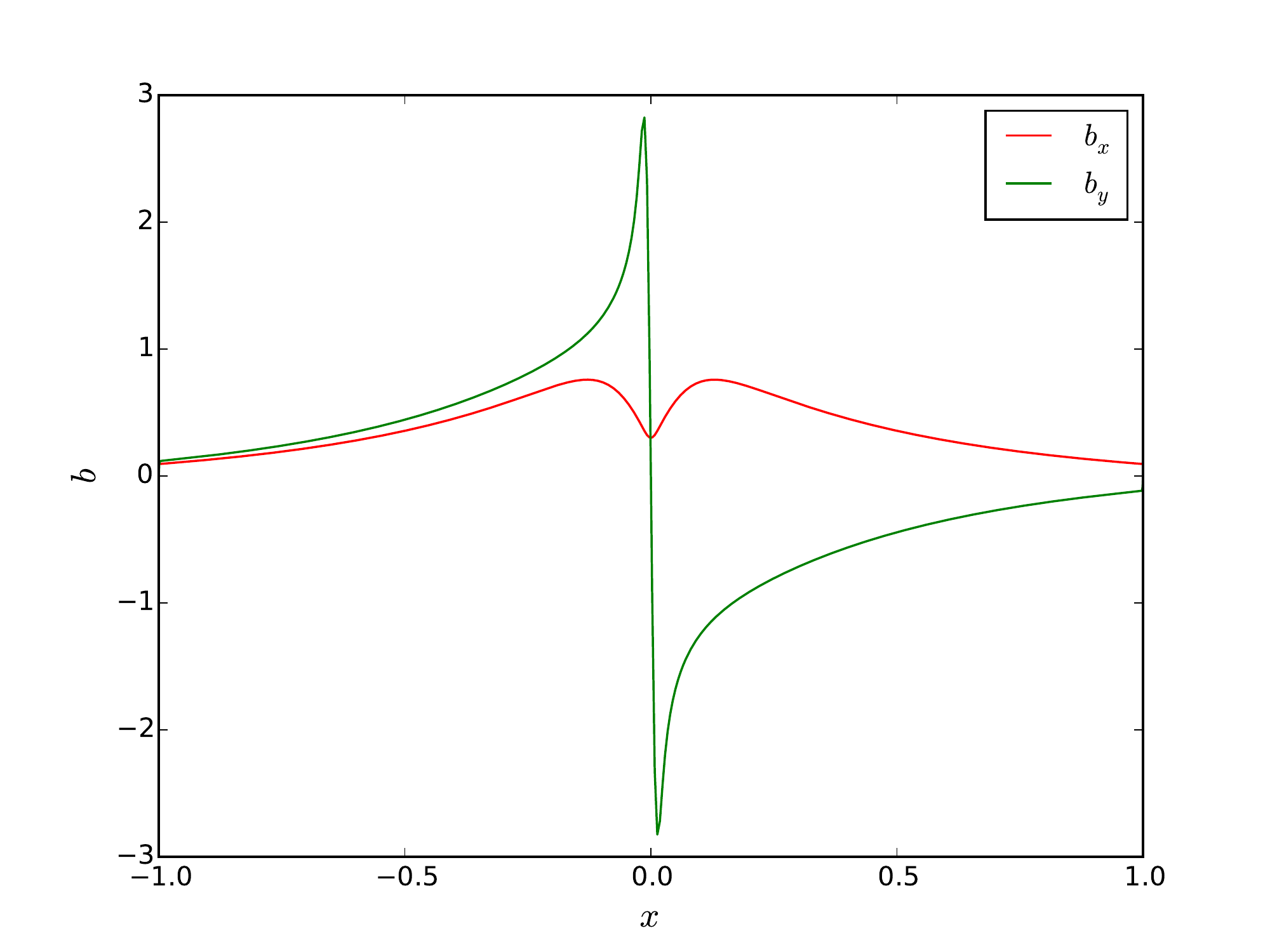}
\caption{The eigenfunctions $b_{x}$ and $b_{y}$ for the fastest growing mode for the case Z01-10. }
\label{Fig:4}
\end{figure}

The inclusion of $B_{y}$ has a mild stabilising effect, reducing the growth rate for given wave number and pushing the maximum growth rate to a higher wave number, as shown in Fig.~\ref{Fig:3} where the dispersion relation is plotted. The eigenfunctions $b_{x}$ and $b_{y}$ for the fastest growing mode with parameters $x_{0}=0.1$ and $R_{H}=1000$  are plotted in Fig.~\ref{Fig:4}, showing that the fastest growing eigenmode consists of oppositely directing $b_{y}$ components on either side of the current sheet and a $b_{x}$ component  with a local minimum at $x=0$. 

\begin{table}
\caption{Summary of the linear stability calculation for the runs with $B_{z,0}=1$, $B_{y,0}=0$ and $k_{y}=0$. The first column is the name of the run, the second the thickness of the reversal area $x_{0}$, the third the value of $B_{y,0}$, the fourth the Hall parameter $R_{H}$, the fifth the wave number at which the maximum growth rate occurs, and the sixth the maximum value of the growth rate. } 
\centering
\begin{tabular}{l c c r r l} \hline\hline
NAME & $x_{0}$ &$B_{y,0}$& $R_{H}$ & $k_{z}$ &$\gamma$\\ \hline
Z05-1 &0.5& 0&  100 &	0.781	& 0.218\\
Z05-2 &0.5&0&200&	0.741	&0.189\\
Z05-4&0.5&0&400	&0.694	&0.156\\
Z05-6&0.5&0&600	&0.663&	0.138\\
Z05-10&0.5&0&1000&	0.623	&0.117\\
Z05-15 &0.5&0&1500&	0.589	&0.102\\
Z05-20&0.5&0& 2000&	0.567	&0.0925\\
\hline
Z02-1 &0.2& 0&  100 &	1.894	&1.525\\
Z02-2 &0.2&0&200&	1.785	&1.315\\
Z02-4&0.2&0&400	&1.652	& 1.095\\
Z02-6&0.2&0&600	&1.565	& 0.973\\
Z02-10&0.2&0&1000&	1.456&	0.830\\
Z02-15 &0.2&0&1500&1.372	& 0.728\\
Z02-20&0.2&0& 2000&1.319	& 0.661\\
\hline
Z01-1 &0.1& 0&  100 &	3.757&	6.150\\
Z01-2 &0.1&0&200&	3.530	&5.306\\
Z01-4&0.1&0&400	&3.255	&4.426\\
Z01-6&0.1&0&600	&3.086	&3.937\\
Z01-10&0.1&0&1000&2.866	&3.368\\
Z01-15 &0.1&0&1500&2.692	&2.959\\
Z01-20&0.1&0& 2000&2.564	&2.693\\
\hline  \hline
\end{tabular}
\label{Table:1}
\end{table}
\begin{table}
\caption{Summary of the linear stability calculation for the runs with $B_{z,0}=1$, $B_{y,0}=1$ and $k_{y}=0$. The columns are as in Table 1.} 
\centering
\begin{tabular}{l c c r r l} \hline\hline
NAME & $x_{0}$ &$B_{y,0}$& $R_{H}$ & $k_{z}$ &$\gamma$\\ \hline
Y05-1 &0.5& 1&  100 &	0.884	&0.178\\
Y05-2 &0.5&1&200&0.851	&0.158\\
Y05-4&0.5&1&400	&0.810	&0.132\\
Y05-6&0.5&1&600	&0.784	&0.117\\
Y05-10&0.5&1&1000&0.752	&0.0988\\
Y05-15 &0.5&1&1500&	0.728&	0.0856\\
Y05-20&0.5&1& 2000&	0.711&	0.0770\\
\hline
Y02-1 &0.2& 1&  100 &	2.067	&1.352\\
Y02-2 &0.2&1&200&	1.958	&1.194\\
Y02-4&0.2&1&400	&1.825	&1.008\\
Y02-6&0.2&1&600	&1.735	&0.901\\
Y02-10&0.2&1&1000&1.628	&0.773\\
Y02-15 &0.2&1&1500&1.539	&0.680\\
Y02-20&0.2&1& 2000&1.476	&0.619\\
\hline
Y01-1 &0.1& 1&  100 &	4.110&	5.433\\
Y01-2 &0.1&1&200&3.893	&4.800\\
Y01-4&0.1&1&400	&3.617	&4.061\\
Y01-6&0.1&1&600	&3.440	&3.632\\
Y01-10&0.1&1&1000&3.209	&3.123\\
Y01-15 &0.1&1&1500&3.022	&2.751\\
Y01-20&0.1&1& 2000&2.902	&2.507\\
\hline  \hline
\end{tabular}
\label{Table:2}
\end{table}

\subsection{Non-linear evolution}

Following the rapid exponential growth of the instability and once the perturbing field becomes comparable to the background one, the instability evolves non-linearly. Furthermore, the background field evolves as well, given the  dissipation in the current sheet. Given these limitations that cannot be assessed by the linear model, we explore the full non-linear evolution of the plane-parallel problem. We integrate numerically the full non-linear equation (\ref{HALL}) using a second order Runge-Kutta scheme for the temporal evolution and a second order finite difference scheme for the spatial derivatives. We assumed vacuum boundary conditions in $x$ and periodicity in $z$. The computational domain extends to $\pm1$ in $x$ and to $\pm2$ in $z$. The resolution used for the majority of the runs was $200\times 400$ points in $x$ and $z$, and was tested against higher resolution for some particular cases with good agreement.

We explore a variety of magnetic field configurations. As initial condition, we used the background field given in equation (\ref{BG2}) superimposed with a
small perturbation in the y component, containing up to $2\times10^{-5}$ of the total energy, so that it would trigger any instability. We used configurations of current sheet initial thickness $x_{0}=0.1$ and $x_{0}=0.2$ combining with Hall parameters $R_{H}=200$ and $R_{H}=400$, corresponding to Z01-2, Z01-4, Z02-2, Z02-4, Y01-2, Y01-4 ,Y02-2 and Y02-4 (Tables \ref{Table:1} and \ref{Table:2}). According to the linear calculation, the wavelength of the fastest growing mode corresponding to these backgrounds is smaller than the domain's extent in $z$. In all runs, except when the Y02-2 initial condition was used, we noticed a growth of the perturbation and the formation of the island pattern of the tearing mode. In what follows we will discuss in detail the results of runs with initial conditions Z01-2 and Z01-4 which encapsulate the basic behaviour of the tearing instability. The Ohmic decay of the background field did not allow enough time for the growth of the instability in the case of Y02-2.

\begin{figure*}
\includegraphics[width=0.65\columnwidth]{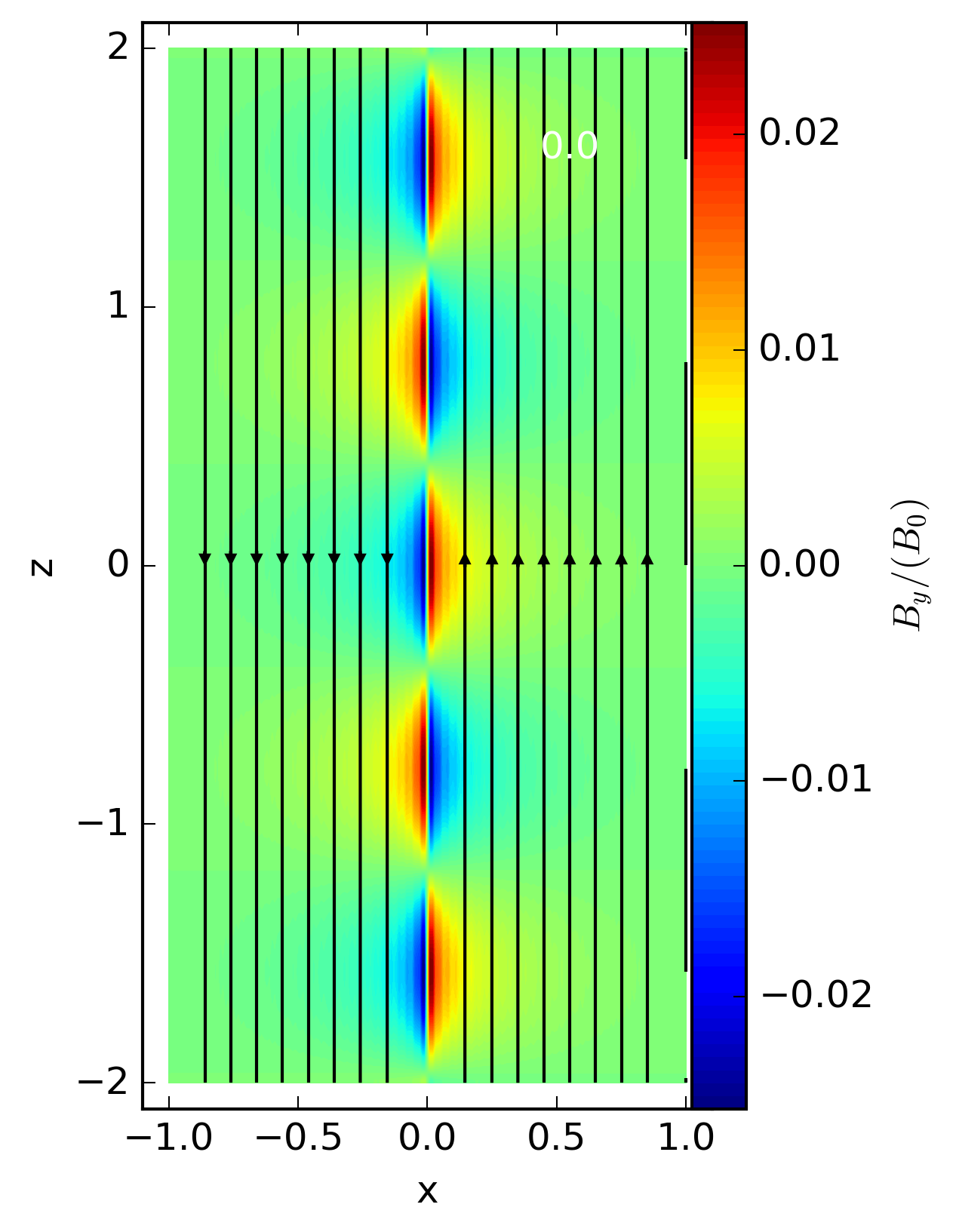}
\includegraphics[width=0.65\columnwidth]{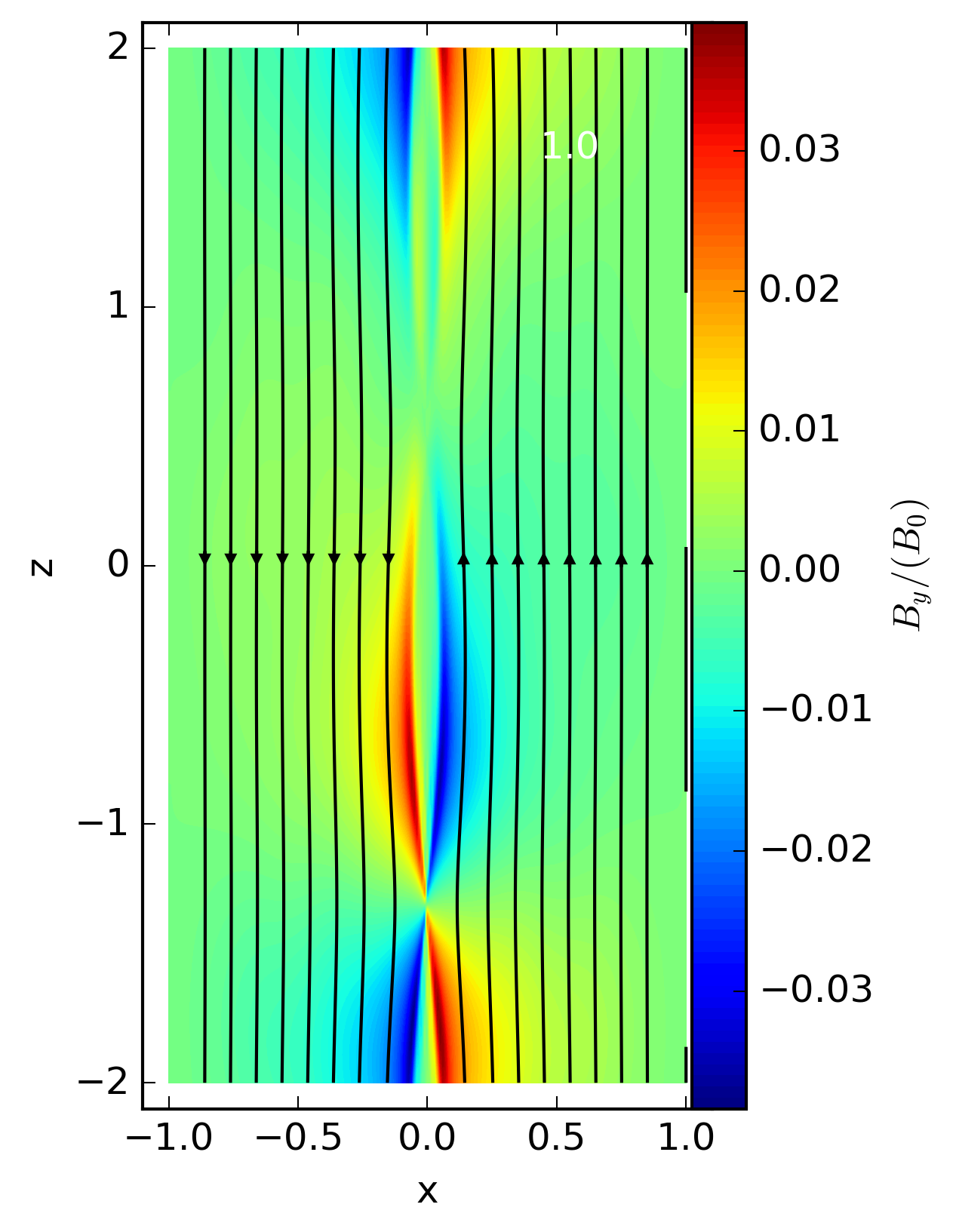}
\includegraphics[width=0.65\columnwidth]{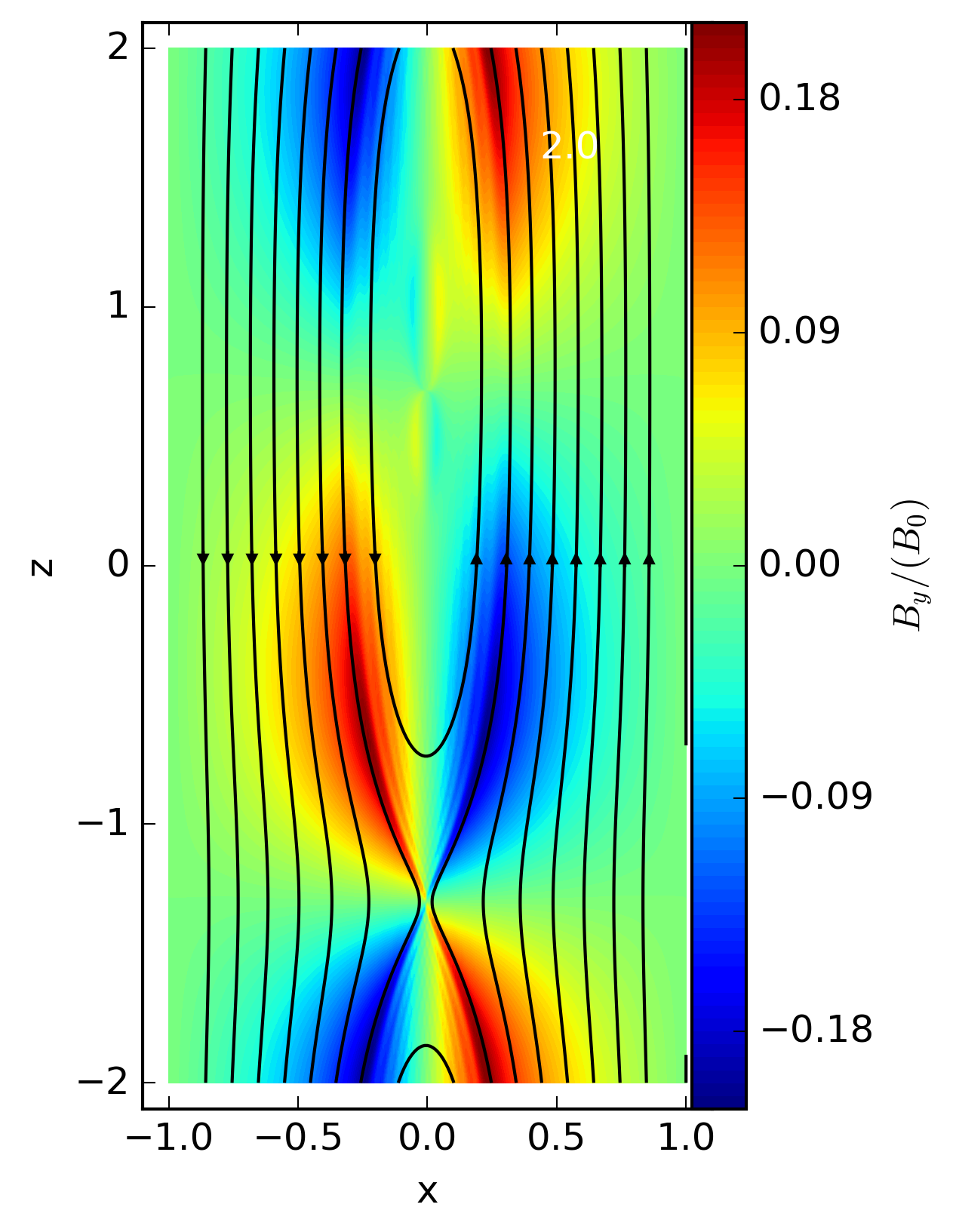}
\caption{The magnetic field for the run using the initial conditions Z01-4  with a superimposed small perturbation in $b_{y}$, at time $t=0$ (left), $\tau_{H}$ (middle) and $2\tau_{H}$ (right), the black lines correspond to the $B_{x}$ and $B_{z}$ components of the field, and the $B_{y}$ component is shown in colour. The magnetic field forms the characteristic islands in the location of the current sheet. As the system evolves and the current sheet decays, the system adopts longer wavelength modes.  }
\label{Fig:5}
\end{figure*}

We plot three snapshots of the magnetic field structure in Fig.~\ref{Fig:5}, at times $t=0$ (left), $t=\tau_{H}$ (middle) and $t=2\tau_{H}$ (right), for the run with initial conditions Z01-4 and some weak perturbation. We find that the strength of the perturbing magnetic field rises from an initial value of $0.02B_{0}$ to $0.18B_{0}$. While the instability is growing, the background field changes as well, in particular the current sheet becomes wider and consequently this has an effect on the growth rates and wavenumbers of the dominant eigenmodes. Thus, the tearing instability is shifted towards longer wave lengths as the wavenumber scales inversely with $x_{0}$. There is also some drift of the newly formed islands along the $z$ direction which is caused by the mixing of modes with different wavelengths and different growth rates. Eventually, once the instability has fully developed it forms the characteristic long-living reconnection islands, right panel of Fig.~\ref{Fig:5}. 

\begin{figure}
\includegraphics[width=\columnwidth]{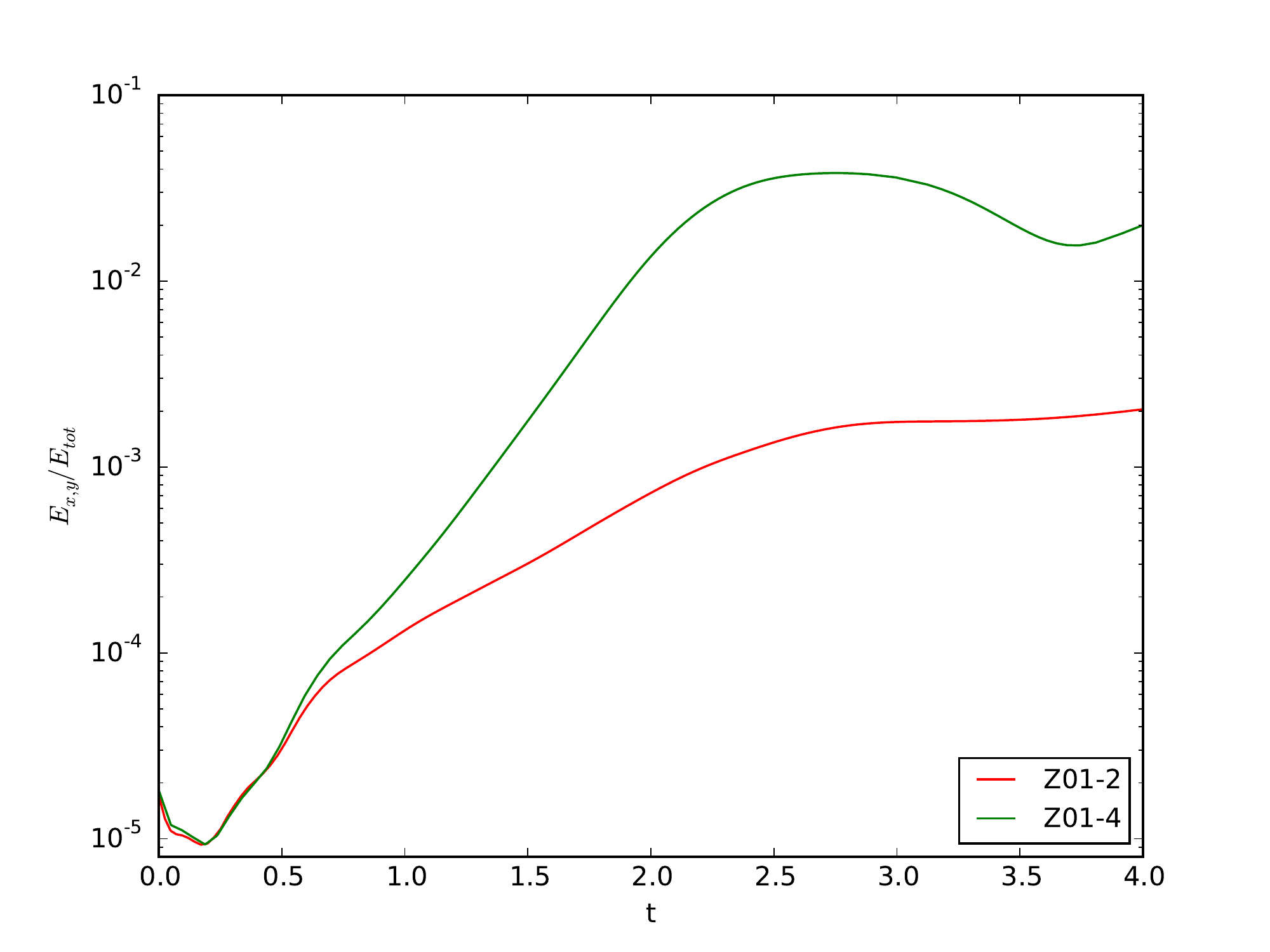}
\caption{The ratio of magnetic energy in the $x$ and $y$ components over the total magnetic energy for two runs with initial conditions that of Z01-2 (red) Z01-4 (green) and a perturbing field containing $2 \times 10^{-5}$ of the total energy. The time is expressed in units of $\tau_{H}$. }
\label{Fig:6}
\end{figure}

\begin{figure}
\includegraphics[width=\columnwidth]{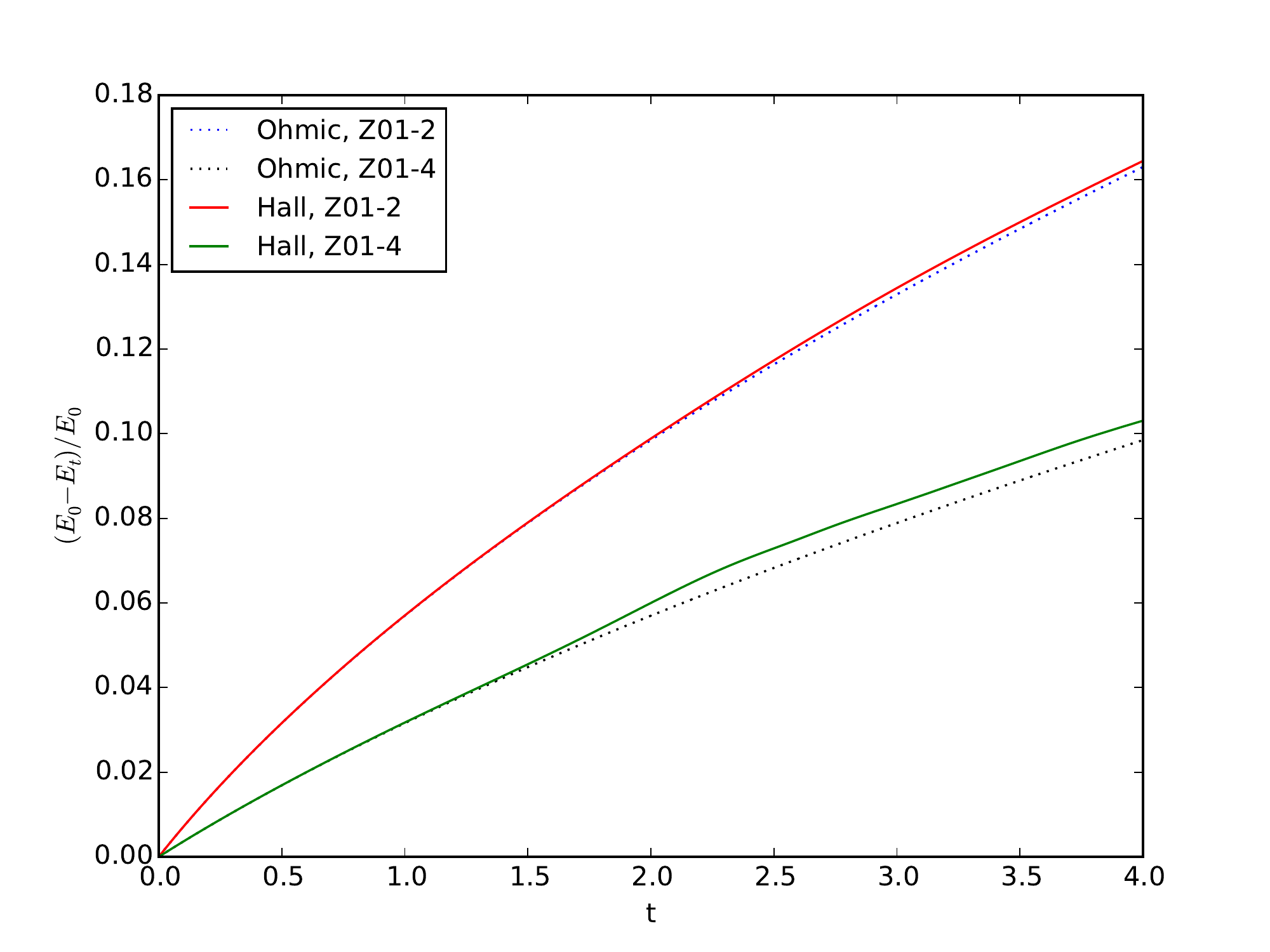}
\caption{The difference of magnetic energy at time  $t$, $E_{t}$ from the initial magnetic energy $E_{0}$, for the runs shown in Fig.~\ref{Fig:6}, solid green and red lines. The same quantity for runs evolving only under the Ohmic dissipation. The decay for the system evolving only under Ohmic dissipation is slower, and the difference is more profound for the higher $R_{H}$. }
\label{Fig:7}
\end{figure}

To probe the instability we used the amount of energy in the $x$ and $y$ components of the magnetic field where we plot the results of two runs with $R_{H}=200$ and $R_{H}=400$, and $x_{0}=0.1$ (initial conditions Z01-2, Z01-4), Fig.~\ref{Fig:6}. Following a short initial transition where energy is dissipated from the perturbation, presumably due to damping of modes with negative growth rates ($t<0.2\tau_{H}$),  we find that the amount of energy in the $x$ and $y$ components rises almost exponentially. This phase lasts until $t=2$ for the $R_{H}=400$ run, and corresponds to a growth rate for the energy $\gamma_{E}=6.8$ implying an approximate growth rate for the amplitude of the perturbation field  $\gamma_{I}\approx \gamma_{E}/2=3.4$. This figure is smaller compared to $4.426$ found in the linear analysis, as expected, since the former takes into account the energy in the various other modes which grow at slower rates, while the latter gives the growth rate of the fastest mode only. The growth of the energy of the run where the Z01-2 initial condition was used saturates earlier and at a lower energy.  At very early times ($0.2<t<0.4\tau_{H}$) the instability at $R_{H}=200$ grows marginally faster than the $R_{H}=400$  due to $\gamma_{Z01-4}<\gamma_{Z01-2}$, however this lasts for a very short time as the background field decays swiftly and widens the magnetic field reversal area. For instance, in a run with $R_{H}=200$ and $x_{0}=0.1$ it takes $\sim 3 \tau_{H}$ for the reversal area to double its size if left to decay Ohmically. This means that the growth rate will drop by a factor of $4$ and the wave number of the fastest growing mode will be multiplied by a factor of $2$. With respect to the energy decay, the inclusion of the instability leads to a faster rate compared to a system evolving solely under Ohmic dissipation with the Hall term switched-off, a result that is more prominent in the case of $R_{H}=400$,  Fig.~\ref{Fig:7}.


\section{Discussion}

Following the description of the linear and non-linear evolution, we conclude that this instability is a resistive tearing mode as it fulfils the criteria set by \cite{Furth:1963}. First it is a resistive instability with a clear dependence on the value of the resistivity, second it appears along the current sheet by breaking up the field lines and third it is a long wavelength instability. We remark though that the physical mechanism between the tearing instability in Electron-MHD and the usual MHD evolution is different. In Electron-MHD, a sole equation for the evolution of the system needs to be solved, equation (\ref{HALL}), whereas in MHD the momentum equation needs to be accounted for, as well. Thus, while in the usual MHD case, the development of the instability results from a sequence of events involving magnetic pressure and tension and plasma pressure, in the Electron-MHD such a description is irrelevant, as the Lorentz force is balanced by the ion lattice and the entire evolution is determined by the magnetic induction equation alone.

\begin{figure}
\includegraphics[width=\columnwidth]{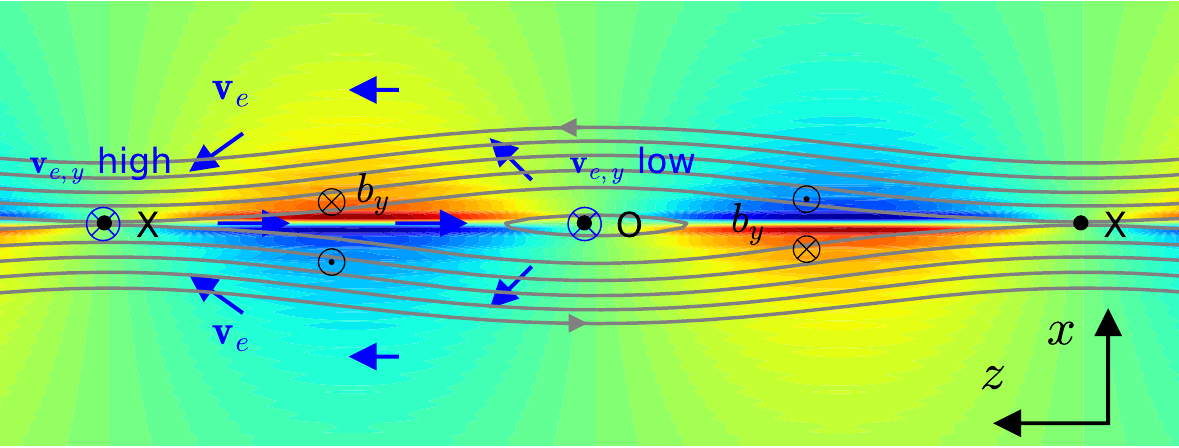}
\caption{Schematic depiction of the instability. We assume a background field directed to $+z$ on the upper half and to the $-z$ on the lower half. The $b_{y}$ component of the perturbation is shown in colour contours with red used to point inwards and blue outwards (also denoted with $\odot$ and $\otimes$ in black). The blue arrows show the electron velocity related to the $b_{y}$ components, and the $\otimes$ blue arrows the electron velocity perpendicular to the plane of the figure. The electron velocity is higher at the $X$ point compared to the $O$ point leading to positive feedback and growth of the $b_{y}$ component. Please refer to the text on the Discussion section for a detailed description of the instability process.}
\label{Fig:8}
\end{figure}

In the Hall-MHD case, the key quantity is the electron fluid velocity advecting the magnetic flux. The electron fluid velocity is uniquely determined by the magnetic field structure through the relation:
\begin{eqnarray}
\bm{v}_{e}=-\frac{c}{4 \pi e n_{e}}\nabla \times \bm{B}\,.
\end{eqnarray}
The instability develops through the steps shown in Fig.~\ref{Fig:8}. The $b_{y}$ component is supported by a current corresponding to the motion of the electron fluid on the plane of the figure with velocity $\bm{v}_{e}$, denoted by blue arrows shown edge-on. Note that since the current is carried by electrons, its flow is antiparallel to the $\bm{v}_{e}$; hereafter we are going to refer to the electron motion to avoid confusion from the oppositely directing current. Considering the $x$ component of the electron flow near the $O$ and $X$ points, we find that it pushes the field lines away from the $O$ point and compresses them towards the $X$ points. Whereas, in the $z$ direction and along the current sheet the electron velocity is from the $X$ point towards the $O$ point. Thanks to resistivity the field lines reconnect at the $X$ point; these newly reconnected field lines shrink around the $O$ point, where they, again due to resistivity, vanish. The compression of the field lines around the $X$ point and the dilution around the $O$ point enhance and suppress the electron flow that runs normal to the plane of the figure, respectively (blue arrow shown tail on). This velocity difference deforms the field lines so that $b_{y}$ is enhanced, closing the positive feedback loop. This is in agreement with the fact that the instability growth rate depends on both the Hall and the Ohmic timescales. The Hall time scale controls the rate at which the field lines move, while the Ohmic times scale set the rate at which the field lines reconnect and essentially controls the supply of magnetic field lines that will move from the $X$ point towards the $O$ point. 

The results of our linear analysis show that the growth rate of the instability is proportional to $R_{H}^{-1/3}$ as opposed to $R_{H}^{-1/5}$ suggested in the analytical approach of \cite{Wood:2014}, while the corresponding wavenumber is proportional to $R_{H}^{-0.15}$ as opposed to $R_{H}^{-1/5}$ suggested there. We find that as long as the boundaries of our calculations are twice as wide compared to the size of field reversal area, their effect on the instability is minimal. These discrepancies are related to the inevitable simplifications made in order to obtain an analytical expression for this instability and the different profiles of the background magnetic field employed not containing a current sheet.

Regarding the full non-linear calculations we find that the instability has a considerable effect on the magnetic field decay once $R_{H}$ is large enough. This is caused by the rapid growth of the initial perturbation and the slow decay of the background state. In the examples simulated we find that for a choice of $R_{H}=400$ the decay rate is clearly enhanced once the instability is close to its saturation point, with milder effect for a choice of $R_{H}=200$. Thus the role of the instability becomes more evident for higher $R_{H}$.

Similar to the variants of the MHD tearing instability, the growth timescale of the E-MHD tearing instability has a mixed dependence of the Hall and the Ohmic  timescales. In the usual MHD tearing instability the growth rate of the tearing instability scales with $\tau_{O}^{-3/5} \tau_{A}^{-2/5}$, where $\tau_{O}$ is the resistive and $\tau_{A}$ the Alfv\`en timescale respectively \citep{Furth:1963}. In relativistic magnetically dominated plasmas the growth rate is the geometric mean of the Alfv\`en timescale and the resistive timescale \citep{Komissarov:2007}. These differences in the growth rates and consequently on the wave numbers reflect the different physical mechanism outlined above.  

The tearing instability in Electron-MHD shares some common properties with the Hall-drift induced magnetic instability which was studied in the linear approximation with uniform \citep{Rheinhardt:2002} and non-uniform \citep{Rheinhardt:2004} background density, and by \cite{Pons:2010} in the non-linear regime. Both instabilities require some non-zero resistivity to operate, as the maximum growth rate of the Hall-drift instability scales as $B_{0}^{q}$, $q<1$, where $B_{0}$ is the magnitude of the magnetic field, thus for negligible resistivity the growth rate becomes zero in physical units. Furthermore both of them are long wavelength instabilities, having positive eigenvalues for $0<k<k_{c}$ where $k_{c}$ is some cut-off wavenumber. They differ on that the Hall-drift instability does not require the presence of a current sheet, even though strong currents are involved, whereas the current sheet is a key element for the development of the tearing instability. Finally, we notice a similarity on the late evolution where the non-linear effects have taken over: in both instabilities the system tends to adopt the longest wavelength permitted by the computational domain leading and the overall dissipation is faster \cite{Pons:2010}.

The role of the Hall effect in the development of the tearing instability has been studied by numerous authors, primarily motivated by experimental results (i.e. \cite{Bodin:1963}). Studies of the effect of Hall current on tearing mode in rotating reverse plasmas of cylindrical geometry have shown that Hall currents combined with rotation of the fluid can suppress tearing modes \citep{Kappraff:1981, Finn:1983, Mirin:1986}. Our approach is different to these ones in two basic aspects. First, we consider the evolution under only Electron-MHD, neglecting other terms arising from Lorentz forces, plasma pressure and inertia, assuming that they are balanced by the elastic forces of the ion lattice, whereas in these works the Hall effect is included as an add-on to normal MHD evolution. Second, the geometry of the system is different assuming a rotating cylinder whereas we study a planar system. Our results are in agreement with those of \cite{Fruchtman:1993}, who showed that the Hall effect can actually lead to a tearing mode in an appropriate planar geometry.

\section{Neutron Star Crust Heating and Outbursts}

Models of global magnetic evolution have shown that a usual outcome of Hall evolution is the development of current sheets \citep{Vainshtein:2000, Hollerbach:2002, Hollerbach:2004, Reisenegger:2007, Geppert:2014, Wood:2015, Gourgouliatos:2016}. Such current sheets are more prominent on the natural boundaries of the crust-core interface \citep{Lander:2013, Beloborodov:2016} and neutron star surface \citep{Thompson:2001b, Lyubarsky:2002}, providing potential sites for the tearing instability.

As shown in the non-linear calculation, a high Hall parameter and a thin layer containing the current sheet are essential for the development of the instability. We can make an estimate of the relative physical parameters using realistic crust models of \citep{Potekhin:1996, Potekhin:1999, Cumming:2004}, where the electron number density at the base of the crust is $\sim 2.5\times 10^{36}$cm$^{-3}$ and the electric conductivity $3.6\times 10^{24}$s$^{-1}$, while we assume that the electron number density at the surface is $ 2.5\times 10^{33}$cm$^{-3}$ and conductivity $3.6\times 10^{22}$s$^{-1}$. Note that the solid surface may extend to lower densities, however at these lower densities the magnetic stresses will be comparable to the breaking stresses of the crust and the assumption of electron MHD does not hold any more \citep{Gourgouliatos:2015a, Lander:2016}. Using the values mentioned above, the Hall parameter at the base of the crust is $R_{H, b}=100B_{15}$ and on the surface $R_{H, s}=1000B_{15}$, where $B_{15}$ is the magnetic field  in units of $10^{15}$G. The Hall timescale at the base of the crust is $\tau_{H,b}=1.5\times 10^{5} B_{15}^{-1}$yr, while on the surface it is $\tau_{H,s}=1.5\times 10^{2} B_{15}^{-1}$yr, where we have assumed a length-scale for the magnetic field $\sim 1$km. Finally we need to get a realistic estimate of the thickness of the current sheet. Numerical simulations place it close to their resolution limit \citep{Hollerbach:2002, Pons:2007,Vigano:2012}, thus in physical dimensions this is $\sim3$m (for a resolution of 346 radial grid points of a $\sim 1$km crust \citep{Vigano:2012} ). Therefore, using these approximations for the quantities appearing in expression (\ref{INST}) the growth rate of the instability near the surface ($\tau_{s}$) and the base of the crust ($\tau_{b}$) of the neutron star is:
\begin{eqnarray}
\tau_{s}\approx 18 ~{\rm days} ~x_{3}^{2} B_{15}^{-2/3} \,, \, \tau_{b}\approx 23 ~{\rm years} ~  x_{3}^{2} B_{15}^{-2/3}
\end{eqnarray}
where $x_{3}$ is the thickness of the current sheet in units of $3$m. Note that the Ohmic decay timescale for a layer of the same thickness close to the surface of a neutron star will be approximately $1.5$year. Assuming that the current sheet covers a fraction $f$ of the surface of the star, whose radius is set to $10$km, the energy that will be contained in this layer will be 
\begin{eqnarray}
E_{I}=1.5\times 10^{44}{\rm ~erg}~B_{15}^{2} x_{3} f\,.
\end{eqnarray}
While thinner layers would lead to a faster growing instability, the instability layer cannot become infinitesimally thin. The release of heat will increase the resistivity of the crust, lower the Hall parameter and eventually dilute the current sheet.  

Release of such amounts of energy in shallow depths have been theorised in order to  explain the bursting behaviour of magnetar outbursts. J1822.3-1606, a low-magnetic field magnetar ($5 \times 10^{13}$G), requires $10^{42}$erg of thermal energy to be deposited between $6\times 10^{8}-6\times 10^{10}$g cm$^{-3}$ \citep{Rea:2012}, or slightly deeper down to $10^{11}$g cm$^{-3}$ \citep{Scholz:2012, Scholz:2014} to power its bursts and subsequent cooling.  Modelling of SGR 0418+5729 has also suggested that a somewhat smaller amount of thermal energy ($10^{41}$erg) in similar depth, is needed to power its bursts \citep{Rea:2013}. In a different magnetar, CXOU J164710.2-455216, whose magnetic field is relatively weak ($<7\times 10^{13}$G), an energy deposition of $\sim 4\times 10^{44}$erg at shallow depths is required to power its bursting events \citep{An:2013}, which could be associated to a much larger part of the crust through a longer wavelength, or alternatively an extremely high magnetic field reaching $10^{16}$G is needed. Finally, in 1E 1048.1-5937, a similar sequence of bursting events has been reported \citep{Archibald:2015} where thermal emission was enhanced in a timeframe of $10^{2}-10^{3}$ days. The energies required by these models can be fulfilled by a current sheet covering as little as $1\%$ of the magnetar surface. We remark that the timescales here are longer than the instantaneous deposition of thermal energy used in cooling models \citep{Pons:2012}, however, for thin current sheets, the generation of Ohmic heat can be as short as few days and will not have a major impact on the post-burst cooling of the magnetar. Another possibility is that this instability operates in conjunction or trigger other types of instabilities suggested to operate in the outer curst, such as the thermoresistive instability \citep{Price:2012} or the thermoplastic instability \citep{Beloborodov:2014, Li:2016}, with the major effect of the tearing mode being on the reduction of the timescales and an increase on the energy efficiency.

Regarding the deeper part of the crust, solutions matching the crustal field to the superconducting core have found that thin current layers naturally form \citep{Henriksson:2013, Lander:2014}, and assuming similar parameters for the thickness of the layer and the strength of the field, the resulting timescale exceeds $\sim 20$ years and cannot be associated to any bursting events. Nevertheless, it may contribute to faster magnetic field decay, affecting the global evolution and quiescent thermal radiation. This effect may be complementary to other processes that have been proposed to operate in the crust-core interface, such as a highly dissipative layer \citep{Pons:2013} and enhance the importance of Hall decay proposed by \cite{Dall'Osso:2012}.

\section{Conclusions}

In this work we have shown that the tearing mode instability operates under the Hall effect and resistivity in the Electron-MHD description. The appearance of the instability is similar to the usual MHD case, developing the characteristic reconnection islands, even though the mechanism is physically different, as the usual concepts of magnetic pressure and tension do not apply in this context. We find that the tearing instability facilitates a faster magnetic field decay, which is more evident for high Hall parameters, without leading to any significant amplification of the strength of the local magnetic field. Considering its role in neutron star magnetic field evolution, we have found it is more likely to occur just below the surface of strongly magnetised neutron stars or close to the crust-core boundary. In the first case the energetics of the instability are consistent with the amount of heat needed for a magnetar burst, which is likely to originate close to the surface, while the associated magnetic field strengths are sufficient to deform the crust. In the latter case, it may provide an extra channel for magnetic field decay and contribute to the quiescent emission. 

We note  that the tearing instability discussed here may be relevant to other systems where evolution under the Hall effect and Electron MHD is important. Namely, the Hall effect is known to operate in protoplanetary discs \citep{Balbus:2001}. \cite{Lesur:2014} showed that the inclusion of ambipolar diffusion and Ohmic decay leads to the formation of magnetic zones and recently, \cite{Bethune:2016} showed that the magnetic field reverses direction within a narrow layer [c.f. Figure 7 of \cite{Bethune:2016}]. We speculate the these reversal regions may be appropriate sites for the development of the tearing instability with implications for the overall evolution of these protoplanetary discs.

Observations of the magnetotail has provided evidence of reconnection activity in the region \citep{Nagai:2001, Runov:2003, Snekvik:2009} and the release of plasmoids due to the Hall effect \citep{Liu:2013}. While the system near the magnetotail is more complicated than the simple Electron-MHD evolution described here, the basic principles described here may be still in operation and enhance the reconnection and the subsequent plasmoid formation.

\section*{Acknowledgments}
We thank our referee Matthias Rheinhardt, whose constructive comments have greatly improved our manuscript. We also thank Andrew Cumming, Paul Schloz for discussion on applications of the tearing instability in neutron star crusts, and Maxim Lyutikov, Serguei Komissarov and Gordon Ogilvie for insightful discussions motivating this work. This work was supported by STFC Grant No. ST/K000853/1.

\bibliographystyle{mnras}
\bibliography{Bibtex.bib}

\begin{thebibliography}{82}
\expandafter\ifx\csname natexlab\endcsname\relax\def\natexlab#1{#1}\fi

\bibitem[{An} et~al.(2013){An}, {Kaspi}, {Archibald} \& {Cumming}]{An:2013}
{An} H., {Kaspi} V.~M., {Archibald} R., {Cumming} A., 2013, \apj, 763, 82

\bibitem[{Archibald} et~al.(2015){Archibald}, {Kaspi}, {Ng}
  et~al.]{Archibald:2015}
{Archibald} R.~F., {Kaspi} V.~M., {Ng} C.-Y., et~al., 2015, \apj, 800, 33

\bibitem[{Aschwanden}(2002)]{Aschwanden:2002}
{Aschwanden} M.~J., 2002, \ssr, 101, 1

\bibitem[{Balbus} \& {Terquem}(2001)]{Balbus:2001}
{Balbus} S.~A., {Terquem} C., 2001, \apj, 552, 235

\bibitem[{Barkov} \& {Komissarov}(2016)]{Barkov:2016}
{Barkov} M.~V., {Komissarov} S.~S., 2016, \mnras, 458, 1939

\bibitem[{Beloborodov} \& {Levin}(2014)]{Beloborodov:2014}
{Beloborodov} A.~M., {Levin} Y., 2014, \apjl, 794, L24

\bibitem[Beloborodov \& Li(2016)]{Beloborodov:2016} Beloborodov, A.~M., \& Li, X.\ 2016, arXiv:1605.09077 

\bibitem[{B{\'e}thune} et~al.(2016){B{\'e}thune}, {Lesur} \&
  {Ferreira}]{Bethune:2016}
{B{\'e}thune} W., {Lesur} G., {Ferreira} J., 2016, \aap, 589, A87

\bibitem[{Biskamp} et~al.(1996){Biskamp}, {Schwarz} \& {Drake}]{Biskamp:1996}
{Biskamp} D., {Schwarz} E., {Drake} J.~F., 1996, Physical Review Letters, 76,
  1264

\bibitem[{Bloch}(1932)]{Bloch:1932}
{Bloch} F., 1932, Zeitschrift fur Physik, 74, 295

\bibitem[{Bodin} \& {Newton}(1963)]{Bodin:1963}
{Bodin} H.~A.~B., {Newton} A.~A., 1963, Physics of Fluids, 6, 1338

\bibitem[Boyd(2001)]{Boyd:2001}
Boyd J., 2001, Chebyshev and Fourier Spectral Methods: Second Revised Edition,
  Dover Books on Mathematics, Dover Publications

\bibitem[{Contopoulos} et~al.(1999){Contopoulos}, {Kazanas} \&
  {Fendt}]{Contopoulos:1999}
{Contopoulos} I., {Kazanas} D., {Fendt} C., 1999, \apj, 511, 351

\bibitem[{Cumming} et~al.(2004){Cumming}, {Arras} \& {Zweibel}]{Cumming:2004}
{Cumming} A., {Arras} P., {Zweibel} E., 2004, \apj, 609, 999

\bibitem[{Dall'Osso} et~al.(2012){Dall'Osso}, {Granot} \&
  {Piran}]{Dall'Osso:2012}
{Dall'Osso} S., {Granot} J., {Piran} T., 2012, \mnras, 422, 2878

\bibitem[{Del Zanna} et~al.(2016){Del Zanna}, {Papini}, {Landi}, {Bugli} \&
  {Bucciantini}]{DelZanna:2016}
{Del Zanna} L., {Papini} E., {Landi} S., {Bugli} M., {Bucciantini} N., 2016,
  \mnras, 460, 3753
  
  \bibitem[Elenbaas et al.(2016)]{Elenbaas:2016} Elenbaas, C., Watts, A.~L., Turolla, R., \& Heyl, J.~S.\ 2016, \mnras, 456, 3282 


\bibitem[{Finn} et~al.(1983){Finn}, {Manheimer} \& {Antonsen}]{Finn:1983}
{Finn} J.~M., {Manheimer} W.~M., {Antonsen} T.~M., 1983, Physics of Fluids, 26,
  962

\bibitem[{Fruchtman} \& {Strauss}(1993)]{Fruchtman:1993}
{Fruchtman} A., {Strauss} H.~R., 1993, Physics of Fluids B, 5, 1408

\bibitem[{Fujisawa} \& {Kisaka}(2014)]{Fujisawa:2014}
{Fujisawa} K., {Kisaka} S., 2014, in { Magnetic Fields throughout Stellar
  Evolution\/}, edited by P.~{Petit}, M.~{Jardine}, H.~C. {Spruit}, vol. 302 of
  { IAU Symposium\/},  427--428

\bibitem[{Furth} et~al.(1963){Furth}, {Killeen} \& {Rosenbluth}]{Furth:1963}
{Furth} H.~P., {Killeen} J., {Rosenbluth} M.~N., 1963, Physics of Fluids, 6,
  459

\bibitem[{Geppert} \& {Vigan{\`o}}(2014)]{Geppert:2014}
{Geppert} U., {Vigan{\`o}} D., 2014, \mnras, 444, 3198

\bibitem[{Goldreich} \& {Reisenegger}(1992)]{Goldreich:1992}
{Goldreich} P., {Reisenegger} A., 1992, \apj, 395, 250


\bibitem[Gourgouliatos \& Cumming(2014)]{Gourgouliatos:2014b} Gourgouliatos, K.~N., \& Cumming, A.\ 2014, Physical Review Letters, 112, 171101 

\bibitem[{Gourgouliatos} \& {Cumming}(2014{\natexlab{b}})]{Gourgouliatos:2014a}
{Gourgouliatos} K.~N., {Cumming} A., 2014{\natexlab{b}}, \mnras, 438, 1618

\bibitem[{Gourgouliatos} \& {Cumming}(2015)]{Gourgouliatos:2015a}
{Gourgouliatos} K.~N., {Cumming} A., 2015, \mnras, 446, 1121

\bibitem[{Gourgouliatos} et~al.(2013){Gourgouliatos}, {Cumming}, {Reisenegger},
  {Armaza}, {Lyutikov} \& {Valdivia}]{Gourgouliatos:2013}
{Gourgouliatos} K.~N., {Cumming} A., {Reisenegger} A., {Armaza} C., {Lyutikov}
  M., {Valdivia} J.~A., 2013, \mnras, 434, 2480

\bibitem[{Gourgouliatos} et~al.(2015){Gourgouliatos}, {Kondi{\'c}}, {Lyutikov}
  \& {Hollerbach}]{Gourgouliatos:2015b}
{Gourgouliatos} K.~N., {Kondi{\'c}} T., {Lyutikov} M., {Hollerbach} R., 2015,
  \mnras, 453, L93

\bibitem[{Gourgouliatos} et~al.(2016){Gourgouliatos}, {Wood} \&
  {Hollerbach}]{Gourgouliatos:2016}
{Gourgouliatos} K.~N., {Wood} T.~S., {Hollerbach} R., 2016, Proceedings of the
  National Academy of Science, 113, 3944

\bibitem[{Henriksson} \& {Wasserman}(2013)]{Henriksson:2013}
{Henriksson} K.~T., {Wasserman} I., 2013, \mnras, 431, 2986

\bibitem[{Hollerbach} \& {R{\"u}diger}(2002)]{Hollerbach:2002}
{Hollerbach} R., {R{\"u}diger} G., 2002, \mnras, 337, 216

\bibitem[{Hollerbach} \& {R{\"u}diger}(2004)]{Hollerbach:2004}
{Hollerbach} R., {R{\"u}diger} G., 2004, \mnras, 347, 1273

\bibitem[{Jones}(1988)]{Jones:1988}
{Jones} P.~B., 1988, \mnras, 233, 875

\bibitem[{Kalapotharakos} \& {Contopoulos}(2009)]{Kalapotharakos:2009}
{Kalapotharakos} C., {Contopoulos} I., 2009, \aap, 496, 495

\bibitem[{Kappraff} et~al.(1981){Kappraff}, {Grossmann} \&
  {Kress}]{Kappraff:1981}
{Kappraff} J., {Grossmann} W., {Kress} M., 1981, Journal of Plasma Physics, 25,
  111

\bibitem[{Kojima} \& {Kisaka}(2012)]{Kojima:2012}
{Kojima} Y., {Kisaka} S., 2012, \mnras, 421, 2722

\bibitem[{Komissarov}(2006)]{Komissarov:2006}
{Komissarov} S.~S., 2006, \mnras, 367, 19

\bibitem[{Komissarov} et~al.(2007){Komissarov}, {Barkov} \&
  {Lyutikov}]{Komissarov:2007}
{Komissarov} S.~S., {Barkov} M., {Lyutikov} M., 2007, \mnras, 374, 415

\bibitem[{Lander}(2013)]{Lander:2013}
{Lander} S.~K., 2013, Physical Review Letters, 110, 7, 071101

\bibitem[{Lander}(2014)]{Lander:2014}
{Lander} S.~K., 2014, \mnras, 437, 424

\bibitem[{Lander}(2016)]{Lander:2016}
{Lander} S.~K., 2016, \apjl, 824, L21

\bibitem[{Landi} et~al.(2015){Landi}, {Del Zanna}, {Papini}, {Pucci} \&
  {Velli}]{Landi:2015}
{Landi} S., {Del Zanna} L., {Papini} E., {Pucci} F., {Velli} M., 2015, \apj,
  806, 131

\bibitem[{Lesur} et~al.(2014){Lesur}, {Kunz} \& {Fromang}]{Lesur:2014}
{Lesur} G., {Kunz} M.~W., {Fromang} S., 2014, \aap, 566, A56

\bibitem[Li et al.(2016)]{Li:2016} Li, X., Levin, Y., \& Beloborodov, A.~M.\ 2016, arXiv:1606.04895 

\bibitem[{Liu} et~al.(2013){Liu}, {Feng}, {Guo} \& {Ye}]{Liu:2013}
{Liu} C., {Feng} X., {Guo} J., {Ye} Y., 2013, Journal of Geophysical Research
  (Space Physics), 118, 2087

\bibitem[{Low}(1973)]{Low:1973}
{Low} B.~C., 1973, \apj, 181, 209

\bibitem[{Lyubarsky} et~al.(2002){Lyubarsky}, {Eichler} \&
  {Thompson}]{Lyubarsky:2002}
{Lyubarsky} Y., {Eichler} D., {Thompson} C., 2002, \apjl, 580, L69

\bibitem[{Marchant} et~al.(2014){Marchant}, {Reisenegger}, {Alejandro Valdivia}
  \& {Hoyos}]{Marchant:2014}
{Marchant} P., {Reisenegger} A., {Alejandro Valdivia} J., {Hoyos} J.~H., 2014,
  \apj, 796, 94

\bibitem[{Mirin} et~al.(1986){Mirin}, {O'Neill}, {Killeen}, {Bonugli} \&
  {Ellis}]{Mirin:1986}
{Mirin} A.~A., {O'Neill} N.~J., {Killeen} J., {Bonugli} R.~J., {Ellis} M.~J.,
  1986, Physics of Fluids, 29, 512

\bibitem[{Nagai} et~al.(2001){Nagai}, {Shinohara}, {Fujimoto}
  et~al.]{Nagai:2001}
{Nagai} T., {Shinohara} I., {Fujimoto} M., et~al., 2001, J. Geophys. Res., 106,
  25929

\bibitem[{Olausen} \& {Kaspi}(2014)]{Olausen:2014}
{Olausen} S.~A., {Kaspi} V.~M., 2014, \apjs, 212, 6

\bibitem[{Pons} \& {Geppert}(2007)]{Pons:2007}
{Pons} J.~A., {Geppert} U., 2007, \aap, 470, 303

\bibitem[{Pons} \& {Geppert}(2010)]{Pons:2010}
{Pons} J.~A., {Geppert} U., 2010, \aap, 513, L12

\bibitem[{Pons} et~al.(2009){Pons}, {Miralles} \& {Geppert}]{Pons:2009}
{Pons} J.~A., {Miralles} J.~A., {Geppert} U., 2009, \aap, 496, 207

\bibitem[{Pons} \& {Rea}(2012)]{Pons:2012}
{Pons} J.~A., {Rea} N., 2012, \apjl, 750, L6

\bibitem[{Pons} et~al.(2013){Pons}, {Vigan{\`o}} \& {Rea}]{Pons:2013}
{Pons} J.~A., {Vigan{\`o}} D., {Rea} N., 2013, Nature Physics, 9, 431

\bibitem[{Potekhin}(1999)]{Potekhin:1999}
{Potekhin} A.~Y., 1999, \aap, 351, 787

\bibitem[{Potekhin} \& {Yakovlev}(1996)]{Potekhin:1996}
{Potekhin} A.~Y., {Yakovlev} D.~G., 1996, \aap, 314, 341

\bibitem[{Price} et~al.(2012){Price}, {Link}, {Epstein} \& {Li}]{Price:2012}
{Price} S., {Link} B., {Epstein} R.~I., {Li} H., 2012, \mnras, 420, 949

\bibitem[{Priest}(1985)]{Priest:1985}
{Priest} E.~R., 1985, Reports on Progress in Physics, 48, 955

\bibitem[{Rea} et~al.(2012){Rea}, {Israel}, {Esposito} et~al.]{Rea:2012}
{Rea} N., {Israel} G.~L., {Esposito} P., et~al., 2012, \apj, 754, 27

\bibitem[{Rea} et~al.(2013){Rea}, {Israel}, {Pons} et~al.]{Rea:2013}
{Rea} N., {Israel} G.~L., {Pons} J.~A., et~al., 2013, \apj, 770, 65

\bibitem[{Reisenegger} et~al.(2007){Reisenegger}, {Benguria}, {Prieto}, {Araya}
  \& {Lai}]{Reisenegger:2007}
{Reisenegger} A., {Benguria} R., {Prieto} J.~P., {Araya} P.~A., {Lai} D., 2007,
  \aap, 472, 233

\bibitem[{Rheinhardt} \& {Geppert}(2002)]{Rheinhardt:2002}
{Rheinhardt} M., {Geppert} U., 2002, Physical Review Letters, 88, 10, 101103

\bibitem[{Rheinhardt} et~al.(2004){Rheinhardt}, {Konenkov} \&
  {Geppert}]{Rheinhardt:2004}
{Rheinhardt} M., {Konenkov} D., {Geppert} U., 2004, \aap, 420, 631

\bibitem[{Rosenbluth} \& {Chang}(1967)]{Rosenbluth:1967}
{Rosenbluth} M.~N., {Chang} D.~B., 1967, Journal of Geophysical Research, 72,
  143

\bibitem[{Runov} et~al.(2003){Runov}, {Nakamura}, {Baumjohann}
  et~al.]{Runov:2003}
{Runov} A., {Nakamura} R., {Baumjohann} W., et~al., 2003, Geophys. Res. Lett.,
  30, 33

\bibitem[{Rutherford}(1973)]{Rutherford:1973}
{Rutherford} P.~H., 1973, Physics of Fluids, 16, 1903

\bibitem[{Scholz} et~al.(2014){Scholz}, {Kaspi} \& {Cumming}]{Scholz:2014}
{Scholz} P., {Kaspi} V.~M., {Cumming} A., 2014, \apj, 786, 62

\bibitem[{Scholz} et~al.(2012){Scholz}, {Ng}, {Livingstone}, {Kaspi}, {Cumming}
  \& {Archibald}]{Scholz:2012}
{Scholz} P., {Ng} C.-Y., {Livingstone} M.~A., {Kaspi} V.~M., {Cumming} A.,
  {Archibald} R.~F., 2012, \apj, 761, 66

\bibitem[{Sironi} \& {Spitkovsky}(2014)]{Sironi:2014}
{Sironi} L., {Spitkovsky} A., 2014, \apjl, 783, L21

\bibitem[{Snekvik} et~al.(2009){Snekvik}, {Juusola}, {{\O}stgaard} \&
  {Amm}]{Snekvik:2009}
{Snekvik} K., {Juusola} L., {{\O}stgaard} N., {Amm} O., 2009, Geophys. Res.
  Lett., 36, L08104

\bibitem[{Somov} \& {Verneta}(1989)]{Somov:1989}
{Somov} B.~V., {Verneta} A.~I., 1989, \solphys, 120, 93

\bibitem[{Spitkovsky}(2006)]{Spitkovsky:2006}
{Spitkovsky} A., 2006, \apjl, 648, L51

\bibitem[{Sturrock}(1966)]{Sturrock:1966}
{Sturrock} P.~A., 1966, \nat, 211, 695

\bibitem[{Thompson} \& {Duncan}(2001)]{Thompson:2001b}
{Thompson} C., {Duncan} R.~C., 2001, \apj, 561, 980

\bibitem[{Uzdensky} \& {Spitkovsky}(2014)]{Uzdensky:2014}
{Uzdensky} D.~A., {Spitkovsky} A., 2014, \apj, 780, 3

\bibitem[{Vainshtein} et~al.(2000){Vainshtein}, {Chitre} \&
  {Olinto}]{Vainshtein:2000}
{Vainshtein} S.~I., {Chitre} S.~M., {Olinto} A.~V., 2000, \pre, 61, 4422

\bibitem[{Vigan{\`o}} et~al.(2012){Vigan{\`o}}, {Pons} \&
  {Miralles}]{Vigano:2012}
{Vigan{\`o}} D., {Pons} J.~A., {Miralles} J.~A., 2012, Computer Physics
  Communications, 183, 2042

\bibitem[{Wareing} \& {Hollerbach}(2009)]{Wareing:2009b}
{Wareing} C.~J., {Hollerbach} R., 2009, Physics of Plasmas, 16, 4, 042307

\bibitem[{Wareing} \& {Hollerbach}(2010)]{Wareing:2010}
{Wareing} C.~J., {Hollerbach} R., 2010, Journal of Plasma Physics, 76, 117

\bibitem[{Wood} \& {Hollerbach}(2015)]{Wood:2015}
{Wood} T.~S., {Hollerbach} R., 2015, Physical Review Letters, 114, 19, 191101

\bibitem[{Wood} et~al.(2014){Wood}, {Hollerbach} \& {Lyutikov}]{Wood:2014}
{Wood} T.~S., {Hollerbach} R., {Lyutikov} M., 2014, Physics of Plasmas, 21, 5,
  052110

\end{thebibliography}

\label{lastpage}

\end{document}